\documentclass[10pt,aps,prd,twocolumn,showpacs,superscriptaddress,eqsecnum,longbibliography,nofootinbib]{revtex4-2}
\usepackage{mathrsfs,amsmath,amsthm,latexsym,amssymb,amsfonts,epsfig,cancel,enumerate,graphicx,subcaption,txfonts}
\usepackage{multirow}
\usepackage{xcolor}
\usepackage[colorlinks, linkcolor=blue, citecolor=blue, urlcolor=blue, plainpages=false, pdfstartview=FitH]{hyperref}
\usepackage{appendix}
\allowdisplaybreaks
\newcommand{\GZU}{School of Physics, Guizhou University, Guiyang 550025, China}

\begin{document}

\title{Shadows and optical appearance of quantum-corrected black holes illuminated by static thin accretions}

\author{Jiawei Chen}
\email{gs.chenjw23@gzu.edu.cn}
\affiliation{\GZU}

\author{Jinsong Yang}
\thanks{Corresponding author}
\email{jsyang@gzu.edu.cn}
\affiliation{\GZU}

\begin{abstract}
Recently, two new quantum-corrected black hole models satisfying covariance have been proposed within the framework of effective quantum gravity. In this paper, we study how the quantum parameter $\zeta$ affects the optical properties of two quantum-corrected black hole models. We first analyze the photon sphere, critical impact parameter, and innermost stable circular orbit as $\zeta$ varies, and constrain $\zeta$ using Event Horizon Telescope data. Additionally, by employing the ray-tracing method to study photon trajectories near the two quantum-corrected black holes, we find that $\zeta$ can reduce the range of impact parameters corresponding to the photon ring and lensed ring. We then examine the optical appearance of these black holes with thin accretion disks, showing $\zeta$ significantly brightens the first model's image but has little effect on the second. Meanwhile, we demonstrate the contributions of the transfer functions to the observed intensity of direct and lensed ring in the observer's field of view, which has rarely been separately illustrated in previous studies. Finally, we study the optical appearance of both quantum-corrected black holes under a static spherical accretion model, with results consistent with the above. Therefore, we conclude that the second quantum-corrected black hole is almost indistinguishable from the Schwarzschild black hole, while the first quantum-corrected black hole can be distinguished from the Schwarzschild black hole through its optical appearance.
\end{abstract}

\maketitle

\section{Introduction}

The successful detection of gravitational waves (GWs) \cite{ LIGO:2017dbh} and the imaging of supermassive black holes (BHs) \cite{EventHorizonTelescope:2019dse} directly validate the predictions of classical general relativity (GR). As one of the most successful theories describing gravity, GR has significantly advanced our understanding of the nature of gravity, especially in terms of spacetime curvature and the propagation of gravitational waves. However, GR has limitations when it comes to describing phenomena under extreme conditions in the universe. For example, GR encounters difficulties when describing singularities within BHs \cite{Penrose:1964wq,Hawking:1970zqf}. At the center of a BH, the curvature of spacetime tends to infinity, leading to the divergence of physical quantities. Additionally, GR cannot effectively describe quantum effects. Therefore, quantum gravity theories have been developed in an attempt to address these problems. The loop quantum gravity (LQG) is one of the most influential theories in this field. LQG directly quantizes spacetime, offering a description that is fundamentally different from GR. Unlike the assumption in classical gravity that spacetime is continuous, LQG posits that spacetime is composed of discrete units \cite{Rovelli:2011eq}. Thus, the structure of spacetime is not continuous but discrete, avoiding the problem of infinite curvature at singularities. The theoretical framework of LQG has a profound impact on modern physics. It not only provides a potential approach to resolving the BH singularity problem but also offers new insights into understanding the initial state of the universe, such as the Big Bang singularity \cite{Bojowald:2008zzb}. Although LQG is still in the stage of theoretical research and lacks direct experimental verification, it presents a highly promising research direction. This background-independent and non-perturbative quantum gravity theory has garnered wide attention, as seen in \cite{Rovelli:1997yv,Rovelli:2011eq,Ashtekar:2005qt,Ashtekar:2004eh,Ashtekar:2013hs,Thiemann:2001gmi,Han:2005km,Modesto:2008im,Perez:2017cmj,Ashtekar:2018lag,Bodendorfer:2019cyv,Kelly:2020uwj,Gan:2020dkb,Sartini:2020ycs,Song:2020arr,Zhang:2020qxw,Zhang:2021wex,Lewandowski:2022zce}.

In the past few decades, extensive research has been conducted on LQG, including the coupling of gravity to matter fields \cite{Lewandowski:2021bkt,Zhang:2022vsl} and the extension of LQG’s quantization methods to higher dimensions \cite{Bodendorfer:2011nx,Han:2013noa,Long:2019nkf,Long:2020wuj,Long:2020agv,Zhang:2011vi,Zhang:2011qq,Zhang:2011vg,Zhang:2011gn,Ma:2011aa,Chen:2018dqz,Zhang:2020smo}. Recently, the conditions for general covariance were rigorously derived for effective quantum gravity within the Hamiltonian framework in \cite{Zhang:2024khj,Zhang:2024ney}. These works resolved potential issues of non-compliance with general covariance in spherically symmetric gravity and resulted in two well-defined BH solutions that depend on quantum parameter $\zeta$. This provides a new platform for studying the quantum effects while preserving covariance. These models have also attracted significant attention \cite{Konoplya:2024lch,Liu:2024soc,Liu:2024wal,Malik:2024nhy,Heidari:2024bkm,Wang:2024iwt,Skvortsova:2024msa,Ban:2024qsa,Du:2024ujg,Lin:2024beb,Konoplya:2025hgp}.

On the other hand, recent breakthroughs in imaging supermassive black holes have provided a crucial window for testing these theories. In BH images, particular attention is given to the photon ring and shadow. When photons approach a BH, their trajectories are significantly influenced by the gravitational field. Some photons cross the event horizon and are absorbed by the BH, corresponding to the central dark region in the observer's view. Others experience minimal deflection and reach distant observers directly, forming the bright regions in the BH image. The boundary between these bright and dark regions, determined by the unstable photon orbits, is known as the BH shadow boundary. For a Schwarzschild BH, the shadow is a perfect circle, whereas for a Kerr BH, the dragging effect causes the shadow’s shape to deviate from circularity. Furthermore, when considering additional physical parameters, the shape of the BH shadow can also be affected. After Synge's pioneering work \cite{Synge:1966okc}, up to now, extensive research has been conducted on BH shadows \cite{Hioki:2009na,Amarilla:2010zq,Abdujabbarov:2016hnw,Tsukamoto:2017fxq,Liu:2020ola,Kumar:2019ohr,Contreras:2019cmf,Jusufi:2020odz,Jha:2023rem,Sanchez:2024sdm}.

However, considering the real physical scenario, the matter around the BH is drawn toward it due to gravitational field, forming an accretion disk \cite{Abramowicz:2011xu}. The image of the BH primarily depends on the physical properties of these accretion flows, the radiative properties of the matter, the geometry of the accretion disk, and the gravitational field surrounding the BH. Different types of accretion flows (such as those based on varying material distributions, temperatures, and rotation speeds) can produce different optical effects. In the study of BH optical appearances, simpler accretion models are often considered, such as the geometrically thin static accretion disk model \cite{Gralla:2019xty} and the spherical accretion model \cite{Bambi:2012tg}. Although these models simplify many physical details, they are effective in describing key phenomena in the BH accretion process and provide optical appearances similar to those observed. Currently, research on BH optical appearances has been conducted within various theoretical frameworks, as seen in \cite{Peng:2020wun,Cardoso:2021sip,Gan:2021xdl,Zeng:2021dlj,Rosa:2022tfv,Wang:2022yvi,Zeng:2022pvb,Yang:2022btw,Huang:2023ilm,daSilva:2023jxa,Wang:2023vcv}.

Therefore, in this paper, we will explore the impact of quantum parameter on BH images through the optical appearances of two quantum-corrected BHs \cite{Zhang:2024khj} under different accretion models within the framework of effective quantum gravity. Specifically, we will use ray-tracing methods to study the photon trajectories and photon rings around two quantum-corrected BHs. We will further present the optical appearances of the two quantum-corrected BH models surrounded by geometrically thin accretion disks and static spherical accretion, and compare them with the Schwarzschild case to examine the effect of quantum parameter on the BH photon rings and optical appearances.

The structure of our article will be organized as follows. In Sec. \ref{section2}, we briefly recall two static spherically symmetric quantum-corrected BHs. Then, we investigate the photon spheres of these BHs and use the observational data from M87* and Sgr A* to constrain the quantum parameter. In Sec. \ref{section3}, within the constraint range of $\zeta$ obtained in Sec. \ref{section2}, we discuss the motion trajectories of photons near the two quantum-corrected BHs. We consider the optical appearance of BHs with different values of $\zeta$ surrounded by static thin accretion disks and compare them with the Schwarzschild case. In Sec. \ref{section4}, we further consider the optical appearance of the two quantum-corrected BHs surrounded by static spherical accretions. Finally, we provide a conclusion in Sec. \ref{section5}. Throughout this paper, we adopt the geometric unit $G=c=1$.

\section{Quantum-Corrected black Hole Spacetimes}\label{section2}

Einstein proposed the principle of general covariance, which states that the mathematical expressions of physical laws remain invariant under any coordinate transformation. In classical GR, it is considered that the physical laws can be expressed as tensor equations, thereby satisfying covariance. However, in quantum gravity theories like LQG, it is unclear whether the gravitational Hamiltonian, after quantum corrections are introduced, can satisfy gauge transformations, and thus whether covariance is preserved \cite{Bojowald:2015zha}. Recently, two effective quantum gravity models that satisfy covariance have been proposed \cite{Zhang:2024khj}. In this section, we will briefly review two covariant quantum corrected BHs recently obtained in effective quantum gravity. We then study the motion of photons near these two quantum-corrected BHs.

The metric of the quantum-corrected BH spacetime is given by the following line element \cite{Zhang:2024khj}
\begin{equation}
	d s^{2}=-f(r) d t^{2}+ \frac{1}{g(r)} d r^{2}+h(r)\left(d \theta^{2}+\sin ^{2} \theta d \phi^{2}\right).\label{xianyuan}
\end{equation}
For the first quantum-corrected BH, denoted as BH-I, the metric functions are given as follows:
\begin{equation}
\begin{split}
f(r)&=1-\frac{2 M}{r}+\frac{\zeta^{2}}{r^{2}}\left(1-\frac{2 M}{r}\right)^{2},\\
g(r)&=1-\frac{2 M}{r}+\frac{\zeta^{2}}{r^{2}}\left(1-\frac{2 M}{r}\right)^{2},\\
h(r)&=r^2.
\end{split}
\end{equation}
As for the second quantum-corrected BH, referred to as BH-II, the metric functions are expressed as follows:
\begin{equation}
	\begin{split}
		f(r)&=1-\frac{2M}{r},\\
		g(r)&=1-\frac{2 M}{r}+\frac{\zeta^{2}}{r^{2}}\left(1-\frac{2 M}{r}\right)^{2},\\
		h(r)&=r^2.
	\end{split}
\end{equation}
The above parameters, $M$ and $\zeta$, represent the ADM mass and the quantum parameter, respectively. Obviously, as $\zeta \rightarrow 0$, both quantum-corrected BHs reduce to the Schwarzschild one.

The metric function $g(r)$ for BH-I and BH-II is the same. Therefore, our analysis of the horizon conditions $g(r) =0$ for BH-I and BH-II yields identical results, namely
\begin{equation}
	\begin{split}
		r_{\rm h} = 2 M.
	\end{split}
\end{equation}
It is evident that the horizons $ r_{\rm h} $ of BH-I and BH-II are independent of the quantum parameter $ \zeta$, remaining unchanged as $ \zeta $ varies.

Next, we consider the equations of motion for photons near the BH. Due to the spherical symmetry of spacetime, we focus, for convenience, on the motion of photons confined to the equatorial plane ($\theta = \pi/2$). We define \cite{Wald:1984bk,Liang:2023bk}
\begin{equation}
\begin{split}
-\kappa :=g_{ab}\left(\frac{\partial }{\partial \tau }\right)^a\left(\frac{\partial }{\partial \tau }\right)^b, \label{keq}
\end{split}
\end{equation}
where $\tau$ represents the affine parameter of a null geodesic, and $\left(\frac{\partial }{\partial \tau }\right)^a$ denotes the tangent vector of the geodesic.

For null geodesics, where $\kappa = 0$, we obtain:
\begin{equation}
\begin{split}
0&=g_{ab}\left(\frac{\partial }{\partial \tau }\right)^a\left(\frac{\partial }{\partial \tau }\right)^b\\
&=-f(r)\left(\frac{d t}{d \tau }\right)^2+\frac{1}{g(r)}\left(\frac{d r}{d \tau }\right)^2+h(r)\left(\frac{d \phi }{d \tau }\right)^2.\label{keq1}
\end{split}
\end{equation}
Here, we have already used the condition $\theta = \pi/2$. In the spacetime \eqref{xianyuan}, there exist two Killing vector fields, the timelike Killing field $\left(\partial/\partial t\right)^{a}$ and the axial Killing filed $\left(\partial/\partial \phi\right)^{a}$, leading to two conserved quantities for null geodesics, the energy $E$ and the angular momentum $L$ \cite{Wald:1984bk,Liang:2023bk}
\begin{equation}
\begin{split}
&E:=-g_{a b}\left(\frac{\partial}{\partial t}\right)^{a}\left(\frac{\partial}{\partial \tau }\right)^{b}=f(r) \frac{d t}{d \tau },\\
&L:=g_{a b}\left(\frac{\partial}{\partial \phi}\right)^{a}\left(\frac{\partial}{\partial \tau }\right)^{b}=h(r) \frac{d \phi}{d \tau }.\label{EL1}
\end{split}
\end{equation}
Substituting Eq. \eqref{EL1} into Eq. \eqref{keq1} and simplifying, we obtain
\begin{equation}
\begin{split}
 \left(\frac{d r}{d \phi }\right)^2 =h(r)^2 \left(\frac{1}{b^2}- V_\text{eff}(r)\right) \frac{g(r)}{f(r)}.\label{ef}
\end{split}
\end{equation}
In obtaining Eq. \eqref{ef}, we used $\frac{dr}{d\tau} = \frac{dr}{d\phi} \frac{d\phi}{d\tau}$. Note that $b$ represents the impact parameter, defined as $b \equiv L/E$, and $V_{\text{eff}}$ is the effective potential of the photon
\begin{equation}
\begin{split}
 V_\text{eff}(r)=\frac{f(r)}{h(r)}.\label{Veff}
\end{split}
\end{equation}

\begin{figure}[htbp]
	\centering
	\includegraphics[width=7.3cm]{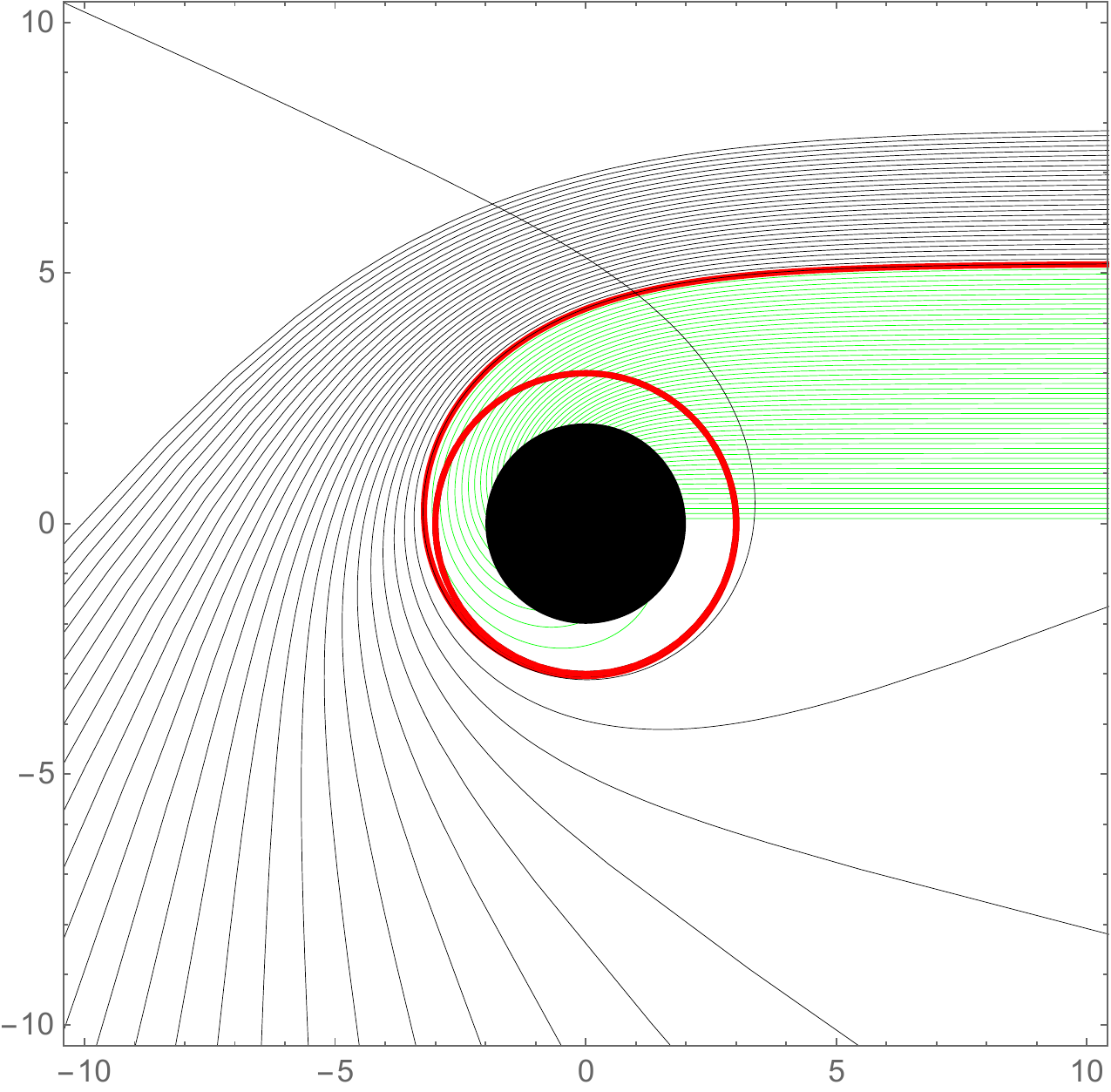}
	\captionsetup{justification=justified, singlelinecheck=false}

	\caption{The figure illustrates three photon trajectories: the green curve represents photons captured by the BH, the red curve corresponds to unstable circular orbits, and the black curve depicts photons scattered by the BH. The black disk represents the BH.}
	\label{fig_photon}
\end{figure}

\begin{figure*}[htbp]
	\centering
	\begin{subfigure}{0.45\textwidth}
		\includegraphics[width=3in, height=5.5in, keepaspectratio]{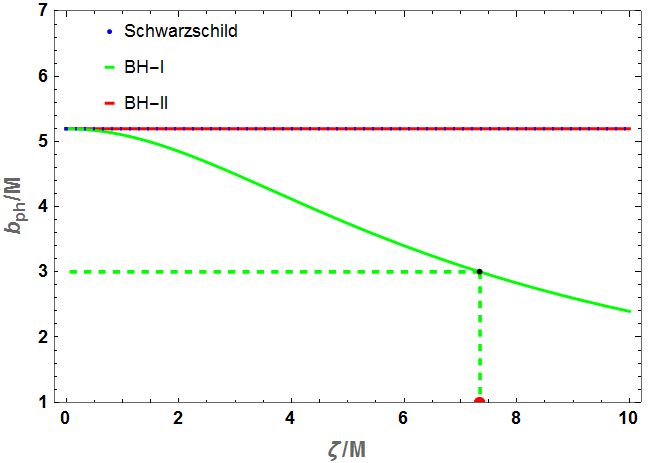}
	\end{subfigure}
	\hfill
	\begin{subfigure}{0.45\textwidth}
		\includegraphics[width=3in, height=5.5in, keepaspectratio]{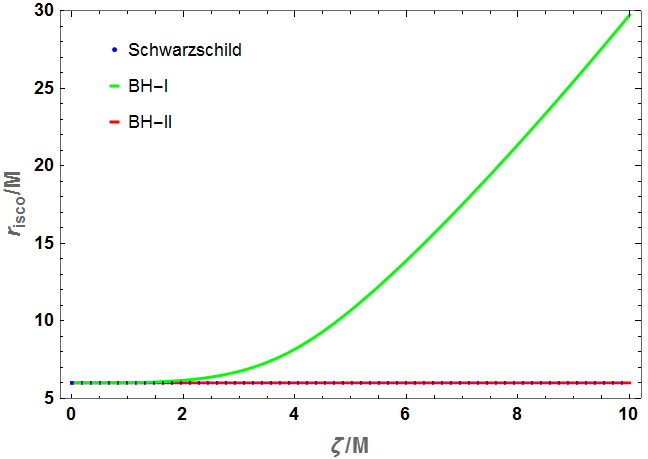}
	\end{subfigure}

	\caption{The figure illustrates the variation of $b_{\rm ph}$ and $r_{\rm isco}$ with $\zeta$ for BH-I and BH-II. The blue dashed line, green solid line, and red solid line represent Schwarzschild, BH-I, and BH-II, respectively.}
	\label{fig:function1}
\end{figure*}

It is clear that the photon’s trajectory is determined by the impact parameter $b$ and the effective potential $V_\text{eff}(r)$. For different values of the impact parameter $b$, there are three possible photon trajectories: one is captured by the BH and falls into it, another is scattered away from the BH to infinity, and a third lies between these two, forming an unstable orbit around the BH. Photons on unstable orbits, when disturbed, either fall into the BH or are scattered to infinity( see Fig.~\ref{fig_photon}). In the observer's view, this corresponds to alternating bright and dark boundaries. The unstable photon sphere orbit is determined by
\begin{equation}
\begin{split}
V_\text{eff}(r) \bigg|_{r = r_\text{ph}}= \frac{1}{b_\text{ph}^2}, \quad
\frac{d V_\text{eff}(r)}{dr} \bigg|_{r = r_\text{ph}} = 0, \quad
\frac{d^2 V_\text{eff}(r)}{dr^2} \bigg|_{r = r_\text{ph}} < 0.\label{Veff1}
\end{split}
\end{equation}
Here, $r_\text{ph}$ is the radius of the photon sphere, and $b_\text{ph}$ is the critical impact parameter corresponding to the photon sphere radius. In general, $b_{\rm ph}$ represents the shadow radius of a static spherically symmetric BH observed by a distant observer. We have:
\begin{equation}
	\begin{split}
	 b_\text{ph}=\frac{r_{\rm ph}}{\sqrt{f(r_{\rm ph})}}\label{bph}.
	\end{split}
\end{equation}
For BH-I, we substitute the metric function into Eq.~\eqref{Veff1} and Eq.~\eqref{bph} to obtain the photon sphere radius $r_\text{ph}$ and the critical impact paramete $b_{\text{ph}}$ as
\begin{equation}
	\begin{split}
		r_{\rm ph}&=3 M,\\
		b_\text{ph}&=\frac{27 M}{\sqrt{27 + \zeta^2 / M^2}}\label{Rbph}.
	\end{split}
\end{equation}
The photon sphere radius $r_{\rm ph}$ remains unchanged with the variation of $\zeta$, while the critical impact parameter $b_{\rm ph}$ exhibits a decreasing trend as $\zeta$ increases. We present these results in Fig.~\ref{fig:function1}. This is reflected in the BH image, indicating that the shadow of BH-I gradually shrinks as $\zeta$ increases. Turning to BH-II, its metric function $f(r)$ is identical to that of the Schwarzschild BH, so its photon sphere radius $r_{\rm ph}$ and critical impact parameter $b_{\rm ph}$ remain $r_{\rm ph}=3 M$ and $b_{\rm ph}=6M$.

When considering the motion of a massive particle, for $\kappa = 1$, by repeating the aforementioned steps, we obtain:
\begin{equation}
\begin{split}
\left(\frac{d r}{d \phi }\right)^2 =h(r)^2 \left(\frac{\bar{E}^2}{\bar{L}^2}- \frac{f(r)}{\bar{L}^2}- \frac{f(r)}{h(r)}\right) \frac{g(r)}{f(r)}\equiv\tilde{V}(r).\label{particleVef}
\end{split}
\end{equation}
Note that here, $\bar{E}$ and $\bar{L}$ represent the energy and angular momentum of the massive particle, respectively. Therefore, the innermost stable circular orbit (isco) of the massive particle is given by
\begin{equation}
\begin{split}
\tilde{V}(r) \bigg|_{r = r_\text{isco}} = 0, \quad
\frac{d\tilde{V}(r)}{dr} \bigg|_{r = r_\text{isco}} = 0, \quad
\frac{d^2\tilde{V}(r)}{dr^2} \bigg|_{r = r_\text{isco}} = 0.
\label{Veff2}
\end{split}
\end{equation}
We perform numerical calculations on the above equations and obtain an expression for $r_{\rm isco}$ in terms of the metric function, namely:
\begin{equation}
	2 g(r_{\rm isco}) \left(-3 + \frac{2 r_{\rm isco} f'(r_{\rm isco})}{f(r_{\rm isco})} - \frac{r_{\rm isco} f''(r_{\rm isco})}{f'(r_{\rm isco})}\right)=0,
\end{equation}
where the prime denotes the derivative with respect to $r$, and $f'(r_{\rm isco})$ represents the derivative evaluated at $r_{\rm isco}$. In the above equation, $g(r_{\rm isco})$ and $f(r_{\rm isco})$ are non-zero, so we obtain the expression for calculating $r_{\rm isco}$,
\begin{equation}
	r_{\rm isco}=\frac{3f(r_{\rm isco})f'(r_{\rm isco})}{2f'(r_{\rm isco})^2-f(r_{\rm isco})f''(r_{\rm isco})}.\label{risco}
\end{equation}
We derive the same expression for calculating $r_{\rm isco}$ as in \cite{Wang:2023vcv} from the more general line element given by \eqref{xianyuan}. Then, we numerically calculate the variation of $r_{\rm isco}$ with $\zeta$ for BH-I and BH-II and present the results in Fig. \ref{fig:function1}.

In Fig. \ref{fig:function1}, the variation of $b_{\rm ph}$ with $\zeta$ is consistent with the previous discussion. The behavior of $r_{\rm isco}$ for BH-II also remains highly consistent with that of the Schwarzschild BH. Regarding BH-I, $r_{\rm isco}$ exhibits a unique dependence, gradually increasing as $\zeta$ increases. From Eqs.~\eqref{Veff1}, \eqref{bph}, and \eqref{risco}, it is evident that, compared to $g(r)$, the metric function $f(r)$ is the primary contributor to $r_{\rm ph}$, $b_{\rm ph}$, and $r_{\rm isco}$. Interestingly, for BH-I, we observe that when $\zeta = 7.35$, as illustrated in Fig.~\ref{fig:function1}, the critical impact parameter satisfies $b_{\rm ph} = r_{\rm ph}$ at this point. This means that photons with the critical impact parameter reach the BH from infinity without any deflection, which is weird. This also reminds us that even when theoretically considering the effects of $\zeta$, its value cannot be chosen arbitrarily. Therefore, we next use observational data from the Event Horizon Telescope (EHT) to constrain the theoretical values of $\zeta$.

\begin{figure*}[htbp]
	\centering
	\begin{subfigure}{0.45\textwidth}
		\includegraphics[width=3in, height=5.5in, keepaspectratio]{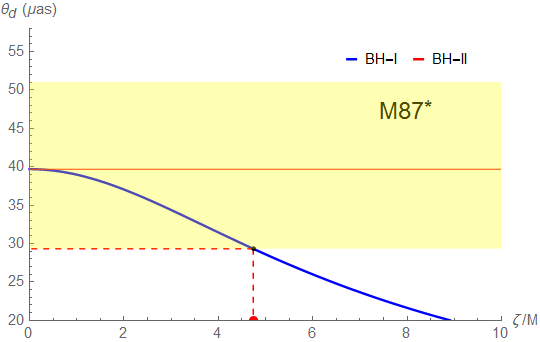}
	\end{subfigure}
	\hfill
	\begin{subfigure}{0.45\textwidth}
		\includegraphics[width=3in, height=5.5in, keepaspectratio]{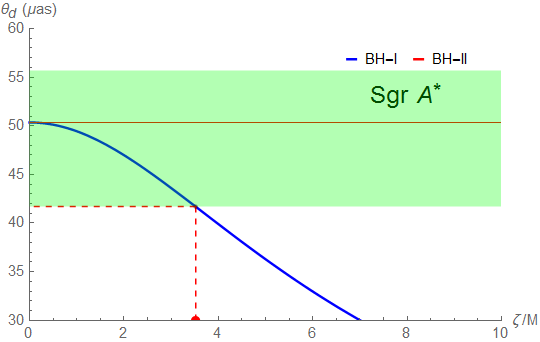}
	\end{subfigure}

	\caption{The figure shows constraints on the quantum parameter $\zeta$ for BH-I and BH-II. The yellow and green regions represent the angular shadow diameters observed by the EHT for M87* and Sgr A*, respectively. The blue and red lines show the variations of the angular diameters of BH-I and BH-II with respect to $\zeta$. BH-II cannot be constrained by observations. For BH-I, under the observational constraints of M87* and Sgr A*, the ranges of $\zeta$ are $0 \leq \frac{\zeta}{M} \leq 4.74$ and $0 \leq \frac{\zeta}{M} \leq 3.52$, respectively.}
	\label{fig:constrain}
\end{figure*}

Existing observational data provide a valuable platform for constraining physically viable BH candidate models. In the context of a static BH, the angular diameter of the BH's shadow can be given by \cite{Kumar:2020owy}
\begin{equation}
	\begin{split}
		\theta_{\rm d} =2 \frac{b_{\text{ph}}}{D},
		\label{thetad}
	\end{split}
\end{equation}
where $b_{\text{ph}}$ is the critical impact parameter of the static BH, and $D$ is the distance from the BH to the observer. The angular shadow radius of M87* is given in \cite{EventHorizonTelescope:2021dqv} as $\theta_{\text{sh}} = 3\sqrt{3}(1 \pm 0.17) \theta_{\rm g}$, where $\theta_{\rm g} = 3.8 \pm 0.4 \, \mu\text{as}$. Therefore, the angular shadow diameter of M87* ranges between 29.32 $\mu\text{as}$ and 51.06 $\mu\text{as}$ \cite{Kuang:2022ojj,Wang:2024lte}. And for Sgr A*, the angular shadow diameter is $48.7 \pm 7 \, \mu\text{as}$ \cite{EventHorizonTelescope:2022wkp}. Different BH masses and distances $D$ lead to different angular shadow diameters. Here, we choose the mass of M87* as $ M = 6.5 \times 10^9 M_{\odot} $ and the distance $ D = 16.8 \, \text{Mpc} $ \cite{EventHorizonTelescope:2019dse}. The mass of Sgr A* is $M = 4.0 \times 10^6 M_{\odot}$, and the distance is $ D = 8.15 \, \text{kpc} $ \cite{EventHorizonTelescope:2022wkp}. Therefore, using Eq. \eqref{bph} and Eq. \eqref{thetad}, we provide the constraint range of the quantum parameter $\zeta$ based on the known angular shadow diameter data in Fig.\ref{fig:constrain}. For BH-I, the upper limit of the quantum parameter is given by $\zeta/M = 4.74$ based on the observational data of M87*. Similarly, the upper limit for Sgr A* is $\zeta/M = 3.52$. As for BH-II, from Figs. \ref{fig:function1} and \ref{fig:constrain}, it is clear that the presence of $\zeta$ does not change the value of the critical impact parameter and angular shadow diameter. Therefore, the $\zeta$ of BH-II cannot be strictly constrained by the known observational data. In the following study, we will ensure that the values of $\zeta$ satisfy these constraint limits.

\section{Rings and images of quantum-corrected BHs with static accretion disks}\label{section3}

\subsection{Photon trajectories around quantum-corrected BHs}

We consider photons with different impact parameters moving towards the equatorial plane of the BH from infinity, and being deflected under the influence of the BH's gravitational field. We adopt the definition of the orbit number from \cite{Gralla:2019xty}, $n =\phi/{(2\pi)}$, to classify the photon trajectories. For $n < 3/4$, it corresponds to direct emission, where the photon passes through the equatorial plane at most once. For $3/4 < n < 5/4$, it corresponds to lensed emission, where the photon passes through the equatorial plane exactly twice. For $n > 5/4$, it corresponds to photon ring emission, where the photon passes through the equatorial plane three times or more.

\begin{figure}[htbp]
	\centering
	\includegraphics[width=8cm]{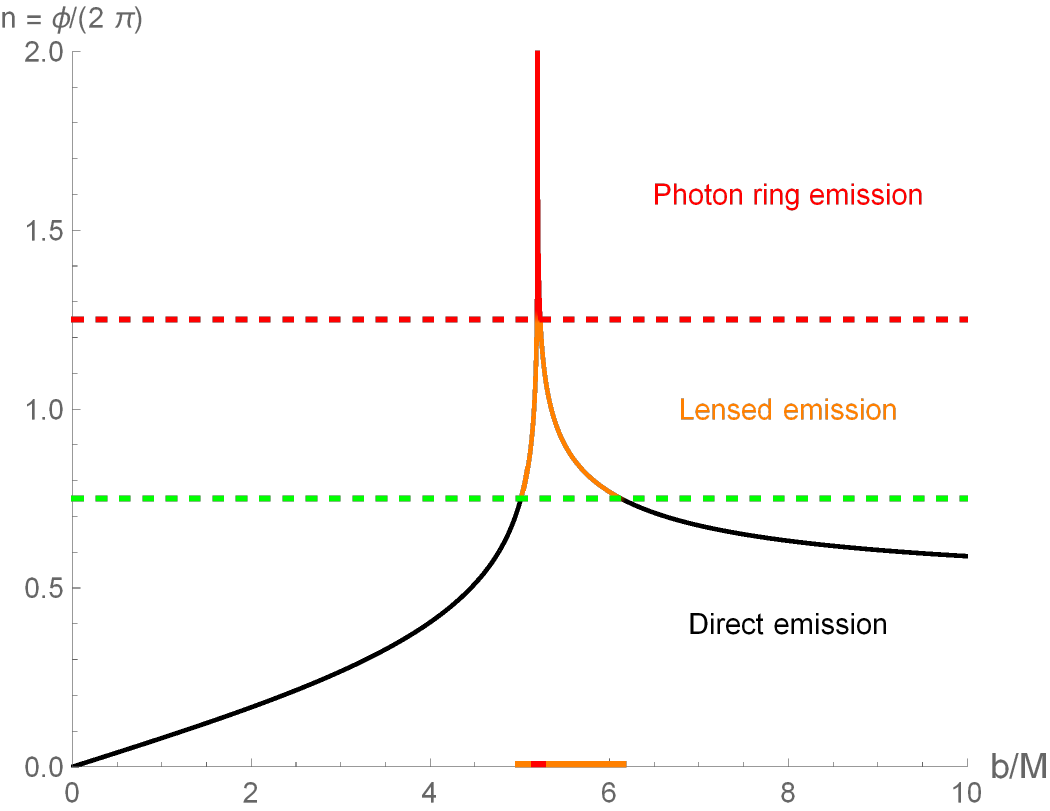}

	\caption{The figure illustrates the relationship between the number of orbits $n$ and $b$ for the Schwarzschild BH, where the red, orange, and black curves represent the photon ring, lensed, and direct emission, respectively.}
	\label{fig_nb}
\end{figure}

\begin{figure*}[htbp]
	\centering
	\begin{subfigure}{0.45\textwidth}
		\includegraphics[width=3.2in, height=6in, keepaspectratio]{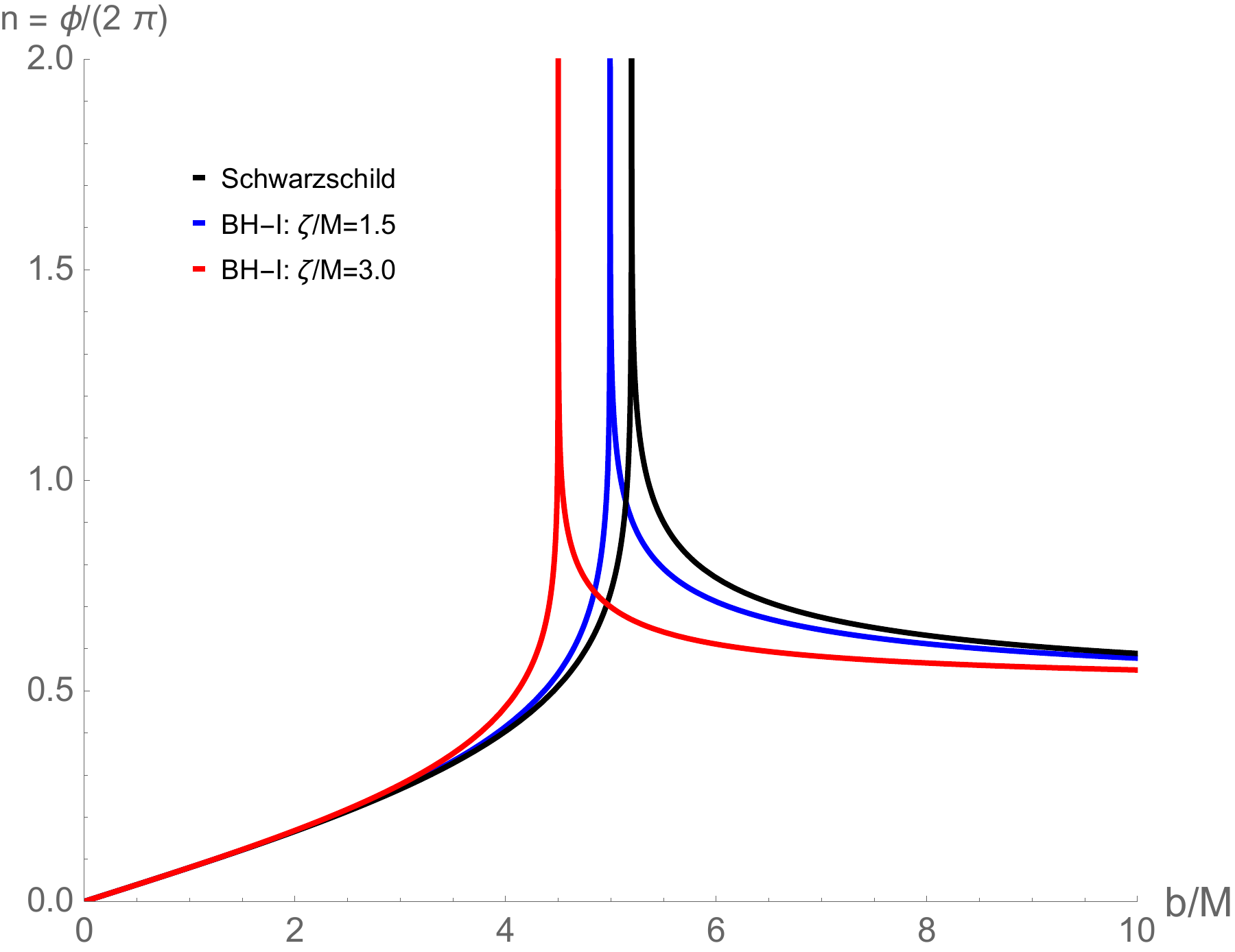}
	\end{subfigure}
	\hfill
	\begin{subfigure}{0.45\textwidth}
		\includegraphics[width=3.2in, height=6in, keepaspectratio]{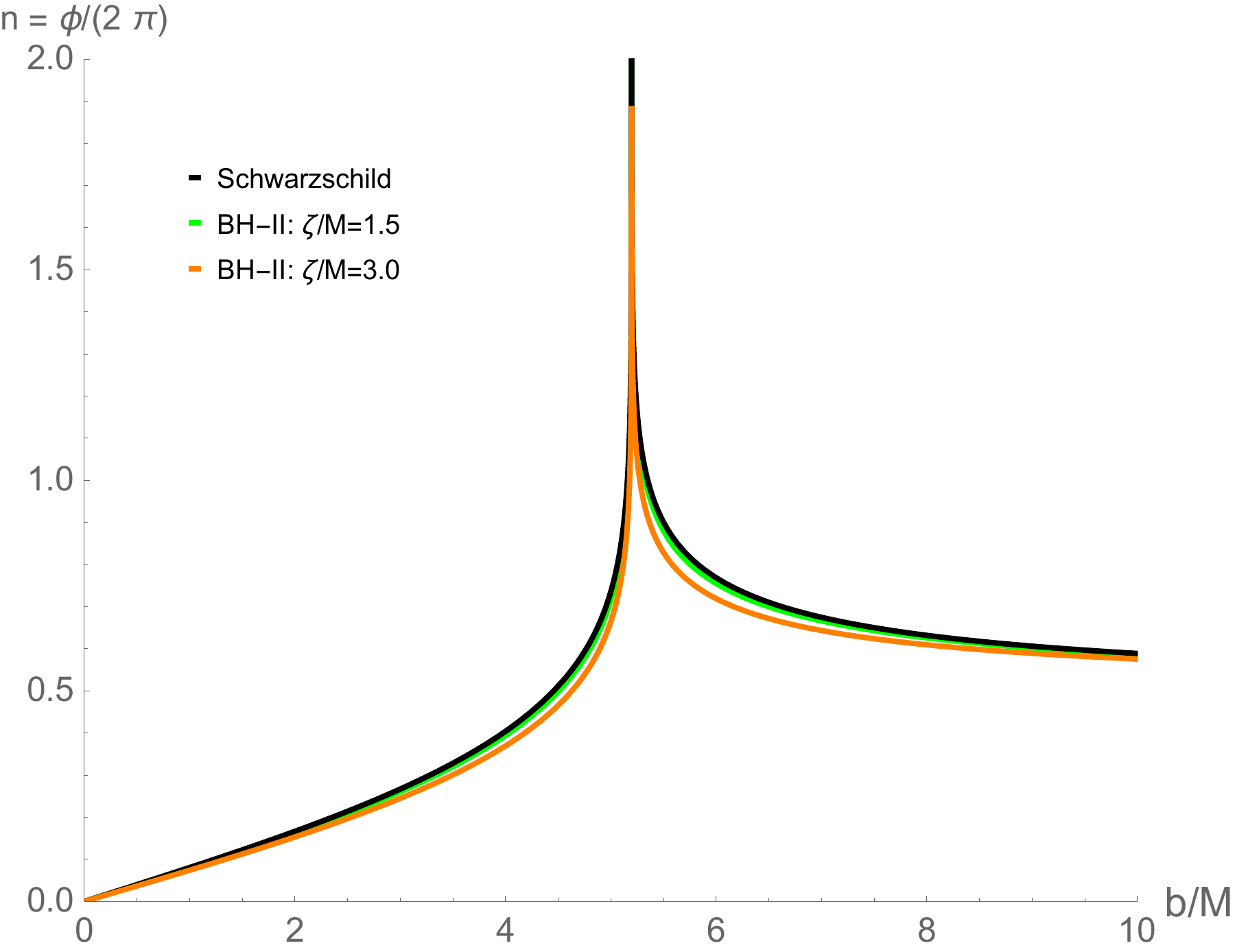}
	\end{subfigure}

	\caption{The left panel shows the total number of orbits $n$ versus the impact parameter $b$ for BH-I at different $\zeta$ values, with red, blue, and black curves representing BH-I ($\zeta=3.0$), BH-I ($\zeta=1.5$), and Schwarzschild, respectively. The right panel displays the same for BH-II, with orange, green, and black curves corresponding to BH-II ($\zeta=3.0$), BH-II ($\zeta=1.5$), and Schwarzschild BH.}
	\label{fig_nb_zong}
\end{figure*}

In the previous section, we derived the equations of motion for photons near the BH and solved them to obtain the photon trajectories. Similarly, the equations of motion determine the relationship between the number of orbits $n$ and the impact parameter $b$. Figure \ref{fig_nb} illustrates this relationship using the Schwarzschild case as an example. In the panel, the red, orange, and black curves represent the photon ring, lensed, and direct emission, respectively. For more details, please refer to \cite{Yang:2022btw}. In addition, we present the total number of orbits $n$ as a function of $b$ for BH-I and BH-II in Fig. \ref{fig_nb_zong}. We observe that their behavior is similar to that of the Schwarzschild case. Therefore, we refrain from extensively displaying the photon trajectory diagrams for BH-I and BH-II here, and instead provide the impact parameter ranges for different emission regions in Table \ref{nb3}.

As illustrated in Fig. \ref{fig_nb_zong} and Table \ref{nb3}, we can clearly observe that after introducing the quantum parameter $\zeta$, the ranges of the impact parameters corresponding to the lensed and photon ring emission regions for both BH-I and BH-II decrease. The larger the value of $\zeta$, the narrower the ranges. However, the difference lies in the fact that for BH-I, the ranges of the impact parameters for the lensed and photon ring regions not only shrink but also shift closer to the BH as $\zeta$ increases. In contrast, for BH-II, the range of the impact parameter for the photon ring region only contracts toward the critical impact parameter $b_{\rm ph}$. The reason for this is that $b_{\rm ph}$ of BH-I decreases as $\zeta$ increases, while $b_{\rm ph}$ of BH-II remains unchanged with variations in $\zeta$.
Moreover, the extent of the shrinkage in the impact parameter range for BH-I is greater than that for BH-II. This demonstrates that the effect of $\zeta$ on BH-II is smaller compared to BH-I.
\begin{table*}[ht]
	\centering
	\resizebox{\textwidth}{!}{
		\begin{tabular}{|c|c|c|c|c|}
			\hline
			\multicolumn{2}{|c|}{\textbf{BHs}} & Direct($n<3/4$) & Lensed($3/4<n<5/4$) & Photon ring($n>5/4$) \\ \hline
			\multicolumn{2}{|c|}{\textbf{Schwarzschild}} & $b<5.01514M$ and $b>6.16757M$ & $5.01514M<b<5.18781M$ and $5.22794M<b<6.16757M$ & $5.18781M<b<5.22794M$ \\ \hline
			\multirow{2}{*}{\textbf{BH-I}} & $\zeta/M=1.5$ & $b<4.85597M$ and $b>5.70478M$ & $4.85597M<b<4.98743M$ and $5.00992M<b<5.70478M$ & $4.98743M<b<5.00992M$ \\ \cline{2-5}
			& $\zeta/M=3.0$ & $b<4.43967M$ and $b>4.7974M$ & $4.43967M<b<4.49891M$ and $4.50398M<b<4.7974M$ & $4.49891M<b<4.50398M$ \\ \hline
			\multirow{2}{*}{\textbf{BH-II}} & $\zeta/M=1.5$ & $b<5.03708M$ and $b>6.03243M$ & $5.03708M<b<5.18954M$ and $5.21965M<b<6.03243M$ & $5.18954M<b<5.21965M$ \\ \cline{2-5}
			& $\zeta/M=3.0$ & $b<5.08643M$ and $b>5.79606M$ & $5.08643M<b<5.19283M$ and $5.20804M<b<5.79606M$ & $5.19283M<b<5.20804M$ \\ \hline
		\end{tabular}
	}

	\caption{The range of impact parameters corresponding to direct, lensed, and photon ring emission for BH-I and BH-II with different values of $\zeta$.}\label{nb3}
\end{table*}

\subsection{Optical appearance of BHs with geometrically thin accretion disk}

Furthermore, we consider a geometrically and optically thin static accretion disk on the BH's equatorial plane (with no extra luminous matter surrounding it), whose emission intensity $I_{\nu_e}^{\rm em}$ depends solely on the radial coordinate $r$. And here $\nu_e$ denotes the emission frequency in a static frame. The intensity received by the observer at the North Pole, corresponding to the emission frequency ${\nu_e}$ from the accretion disk, is given by \cite{Gralla:2019xty}
\begin{equation*}
	I_{\nu_o}^{\rm obs}=g^3 I_{\nu_e}^{\rm em}(r),
\end{equation*}
here $g=\nu_o/\nu_e=\sqrt{f(r)}$ is the redshift factor.

\begin{figure}[htbp]
	\centering
	\includegraphics[width=8cm]{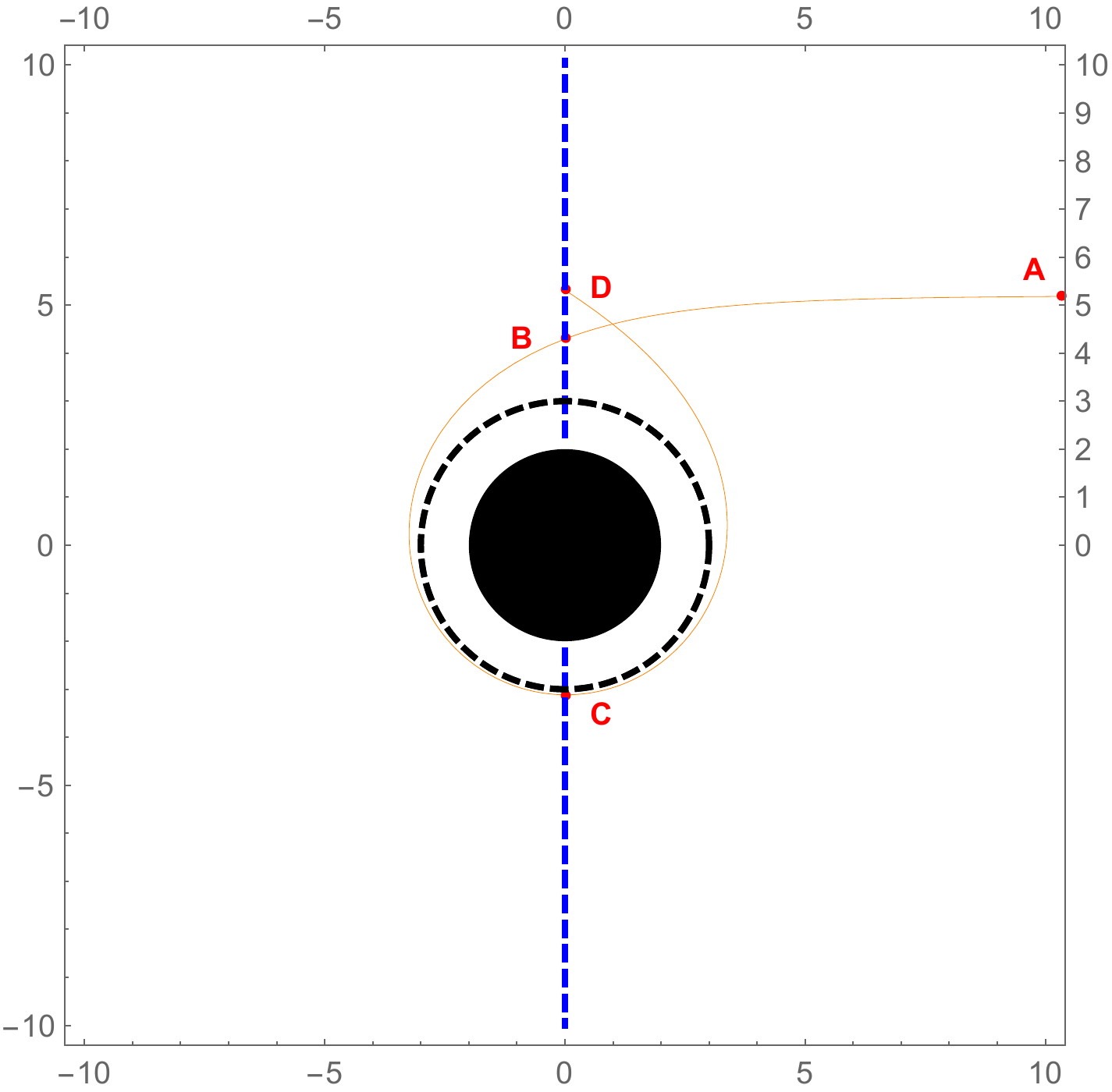}
	\caption{The trajectory of the light rays received by the observer at point $A$ from the BH accretion disk. The blue dashed line, black dashed line, and black disk represent the accretion disk, the photon sphere, and the BH, respectively.}
	\label{fig_yanshi}
\end{figure}

The light emitted from the accretion disk can also be viewed as the backward rays from the observer intersecting the disk. Each intersection corresponds to obtaining brightness from the disk. As shown in Fig. \ref{fig_yanshi}, the rays from point $A$, in addition to intersecting the front of the disk at point $B$, may also deflect around the black hole and intersect with points $C$ and $D$. Therefore, the received intensity at a given impact parameter $b$ is the sum of intensities from each intersection of the rays with the disk. We introduce a transfer function $r_m(b)$ ($m=1,2,3,\cdots$) to describe the radial position of the $m$-th intersection point between a photon with an impact parameter $b$ and the accretion disk of the BH.

\begin{figure}[htbp]
	\centering
	\includegraphics[width=8cm,keepaspectratio]{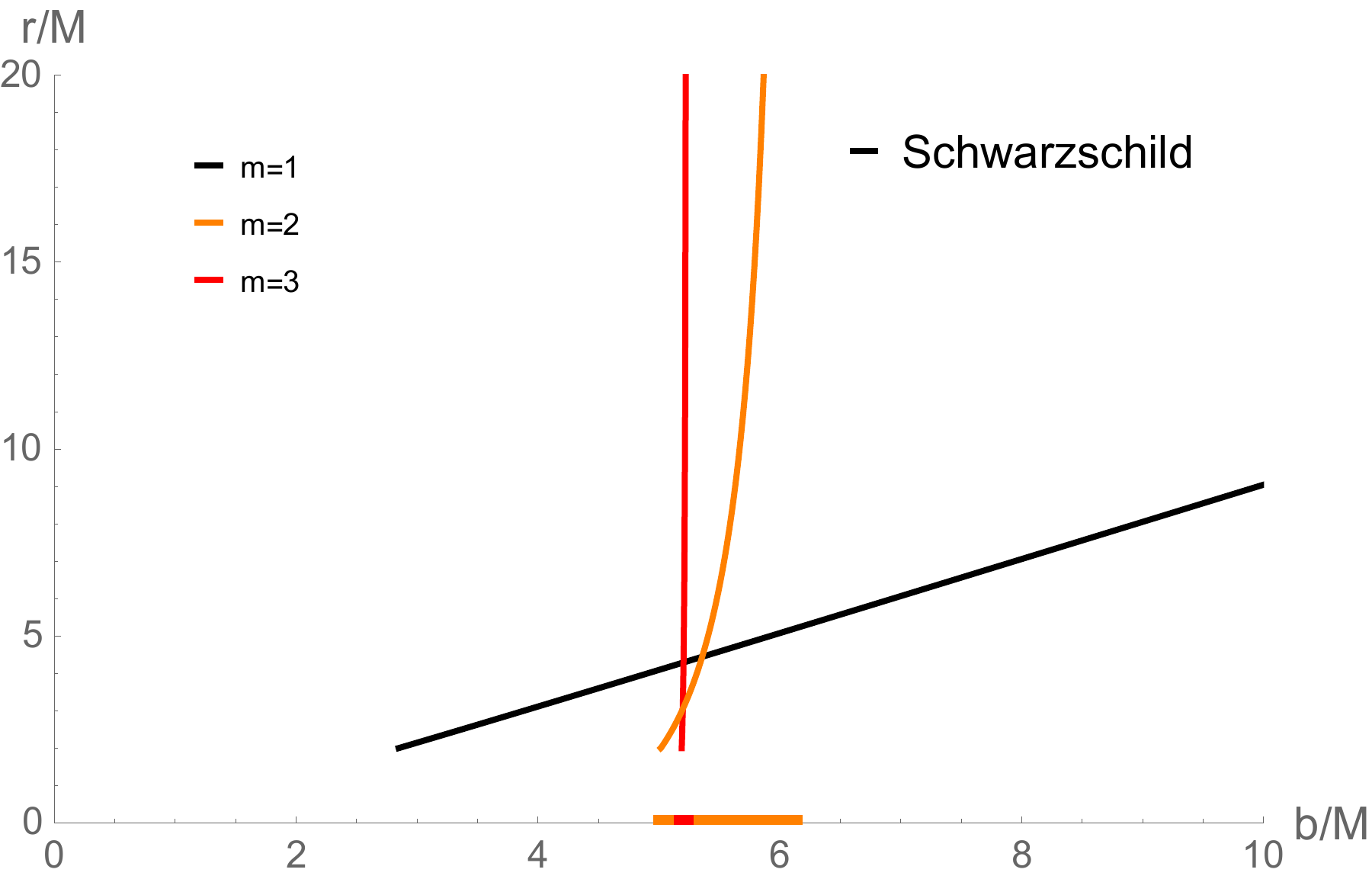}
	\caption{The figure illustrates the variation of the first three transfer functions of the Schwarzschild BH with respect to the impact parameter $b$, where the black, orange, and red curves correspond to $m=1$, $m=2$, and $m=3$, respectively.}
	\label{fig_sch_chuan}
\end{figure}

\begin{figure*}[htbp]
	\centering
	\begin{subfigure}{0.45\textwidth}
		\includegraphics[width=3.2in, height=6in, keepaspectratio]{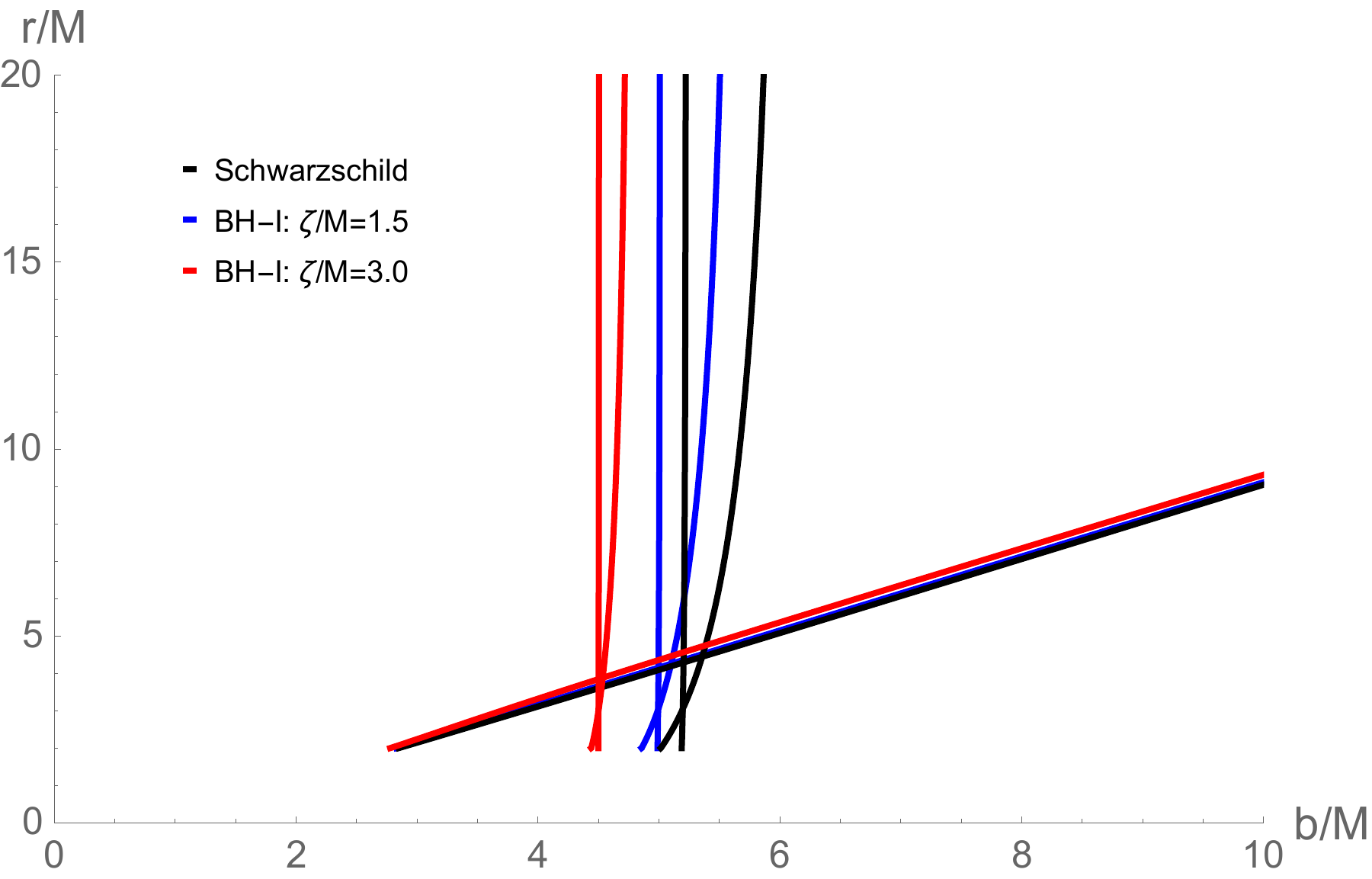}
	\end{subfigure}
	\hfill
	\begin{subfigure}{0.45\textwidth}
		\includegraphics[width=3.2in, height=6in, keepaspectratio]{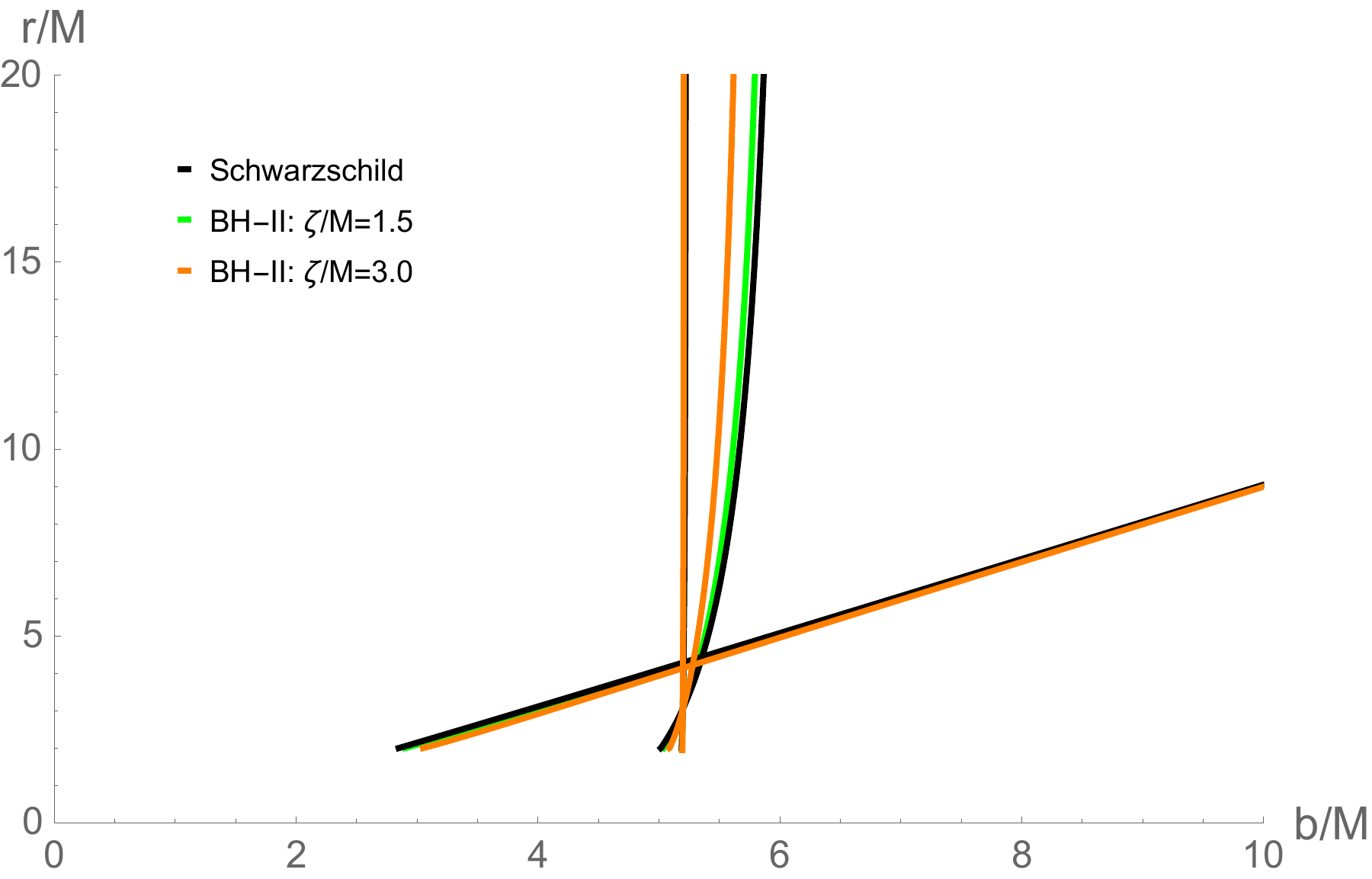}
	\end{subfigure}

	\caption{The left panel shows the transfer functions versus the impact parameter $b$ for BH-I at different $\zeta$ values, with red, blue, and black curves representing BH-I ($\zeta=3.0$), BH-I ($\zeta=1.5$), and Schwarzschild, respectively. The right panel displays the same for BH-II, with orange, green, and black curves corresponding to BH-II ($\zeta=3.0$), BH-II ($\zeta=1.5$), and Schwarzschild BH.}
	\label{fig_chuan_zong}
\end{figure*}

Figure \ref{fig_sch_chuan} illustrates the behavior of the first three transfer functions of Schwarzschild BH, where the black, orange, and red solid lines correspond to $m=1$, $m=2$, and $m=3$, respectively. As shown in the plot, we observe that for the first transfer function ($m=1$), its slope $dr/db$ is nearly 1, indicating that the resulting image originates from direct, lensed, and photon ring emission. For the second transfer function ($m=2$), its slope is much steeper, and the resulting image is primarily from lensed and photon ring emission, with the light from the accretion disk converging to a smaller impact parameter range compared to the first transfer function. For the third transfer function ($m=3$), its slope is extremely steep, and the resulting image is from photon ring emission, with the light from the disk converging to a narrow impact parameter range. The transfer functions of BH-I and BH-II are similar to those of the Schwarzschild BH, and we present them in Fig. \ref{fig_chuan_zong}. We find that the first transfer function, $m=1$, is almost independent of $\zeta$. And $m=2$ indicates that the photon intersects the accretion disk twice, corresponding to the aforementioned lensed emission. Similarly, when $m$ is 3 or greater, it corresponds to the photon ring emission region. Therefore, the behavior of the transfer functions for $m=2$ and $m=3$ as $\zeta$ varies is consistent with the discussion above.

\begin{figure}[htbp]
	\centering
	\includegraphics[width=8cm]{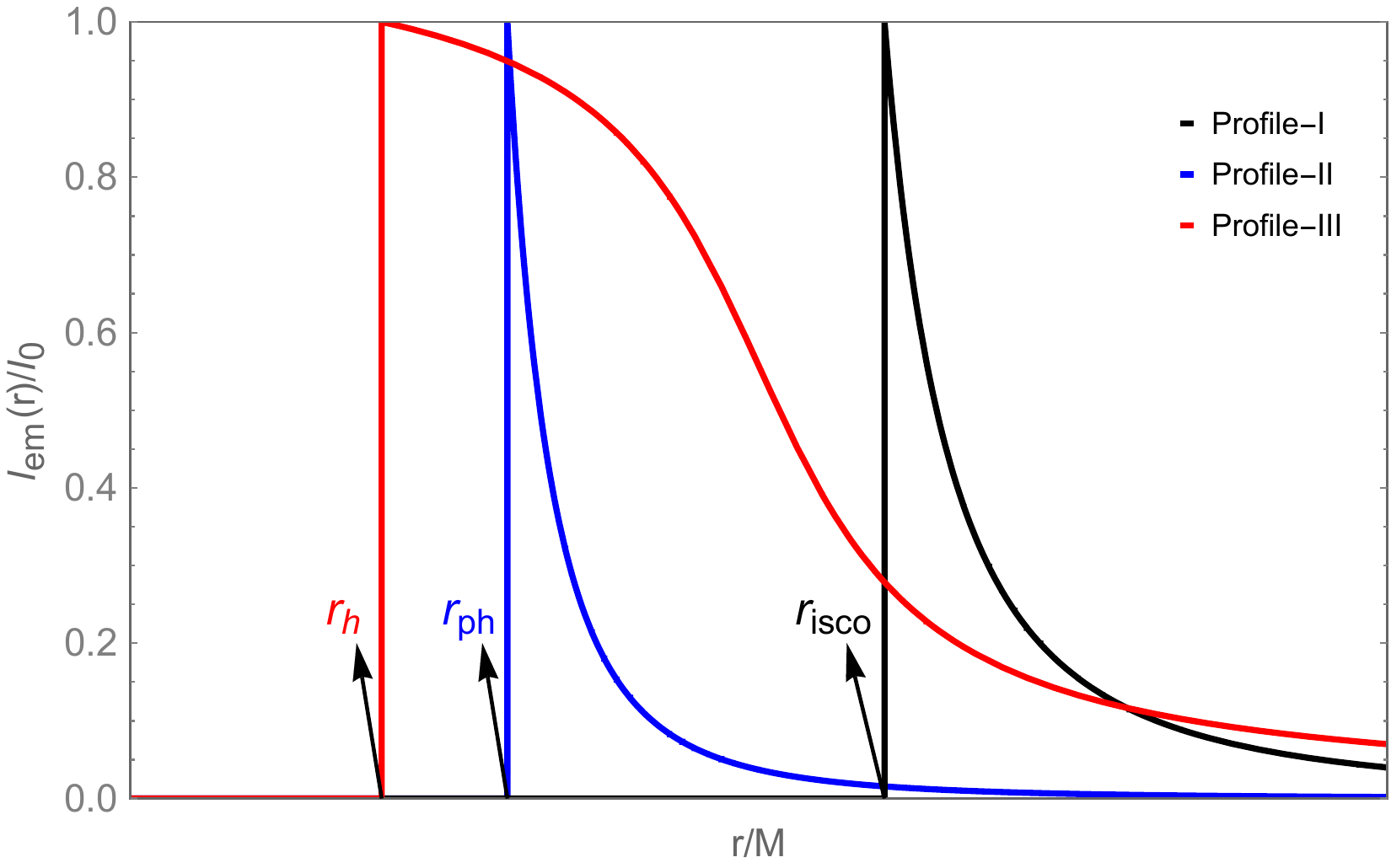}
	\caption{The variation of the emission intensity of the three types of accretion disk models with respect to $r$ is shown.}
	\label{fig_iem}
\end{figure}
Then, the total intensity received by the observer at $b$ can be expressed as:
\begin{equation}
	I_{\rm obs}(b)=\int g^4 I_{\nu_e}^{\rm em}(r) d\nu_e=\sum_{m}f(r)^2I_{\rm em}(r)|_{r=r_m (b)},\label{obs}
\end{equation}
where $I_{\rm em}(r)\equiv \int I_{\nu_e}^{\rm em}(r) d\nu_e$ is the total emitted intensity.
Before obtaining the optical appearance of the quantum-corrected BHs, we need to specify the emission intensity $I_{\text{em}}(r)$ of the accretion disk. Here, we consider three emission models \cite{Gralla:2019xty,Wang:2022yvi}. The first, Profile-I, represents a rapid decay starting from the innermost stable circular orbit $r_{\text{isco}}$. The second, Profile-II, starts decaying from the photon sphere $r_{\text{ph}}$. The third, Profile-III, represents a slow decay starting from the event horizon $r_{\text{h}}$.
\begin{align}
&\text{Profile-I:}\quad I_{\rm em}(r):=
	\begin{cases}
		I_0\left[\frac{1}{r-(r_{\rm isco}-1)}\right]^2, &\hspace{0.8cm} r>r_{\rm isco}\\
		0,&\hspace{0.8cm} r \leqslant r_{\rm isco}
	\end{cases},\label{inten1}\\
&\text{Profile-II:}\quad I_{\rm em}(r):=
	\begin{cases}
		I_0\left[\frac{1}{r-(r_{\rm ph}-1)}\right]^3, &\hspace{0.9cm} r>r_{\rm ph}\\
		0,&\hspace{0.9cm} r\leqslant r_{\rm ph}
	\end{cases},\label{inten2}\\
&\text{Profile-III:}\quad I_{\rm em}(r):=
	\begin{cases}
		I_0\frac{\frac{\pi}{2}-\arctan[r-(r_{\rm isco}-1)]}{\frac{\pi}{2}-\arctan[r_{\rm h}-(r_{\rm isco}-1)]}, &\hspace{0.1cm} r>r_{\rm h}\\
		0,&\hspace{0.1cm} r\leqslant r_{\rm h}
	\end{cases},\label{inten3}
\end{align}
here $I_0$ represents the maximum emission intensity. We then substitute the values of $r_{\text{isco}}$, $r_{\text{ph}}$, and $r_{\text{h}}$ for BH-I and BH-II into Eq. \eqref{inten1}, Eq. \eqref{inten2}, and Eq. \eqref{inten3} to obtain the emission intensity. We show the variation of the emission intensity of the three types of accretion disk profiles with respect to $r$ in Fig. \ref{fig_iem}.

\begin{figure*}[htb]
	\centering
	\begin{subfigure}{0.3\textwidth}
		\includegraphics[height=3cm, keepaspectratio]{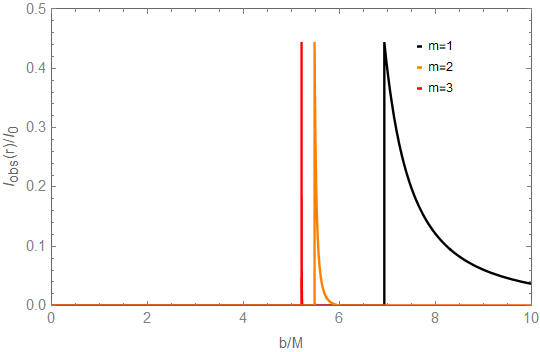}
		\caption{Profile-I}
	\end{subfigure}
	\begin{subfigure}{0.22\textwidth}
		\includegraphics[height=3cm]{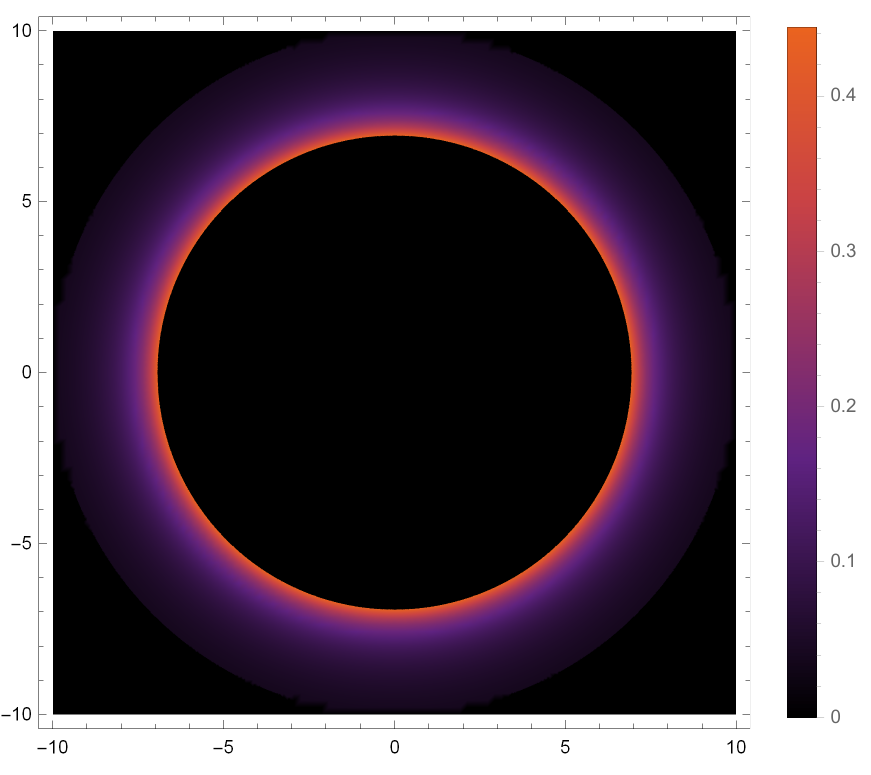}
		\caption{$m=1$}
	\end{subfigure}
	\begin{subfigure}{0.22\textwidth}
		\includegraphics[height=3cm,keepaspectratio]{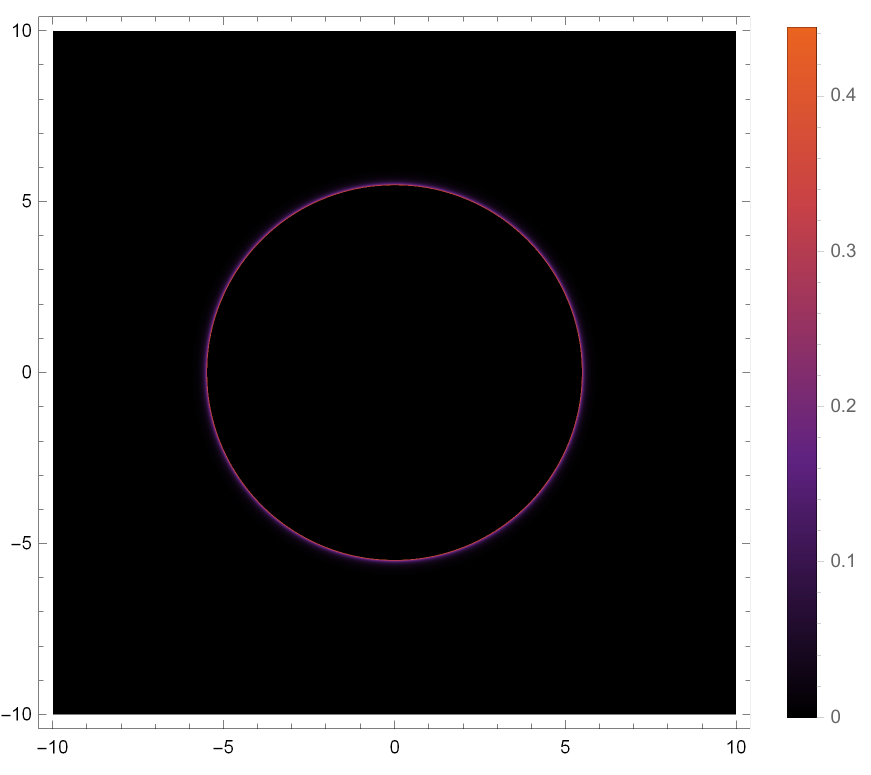}
		\caption{$m=2$}
	\end{subfigure}
 \begin{subfigure}{0.22\textwidth}
 \includegraphics[height=3cm]{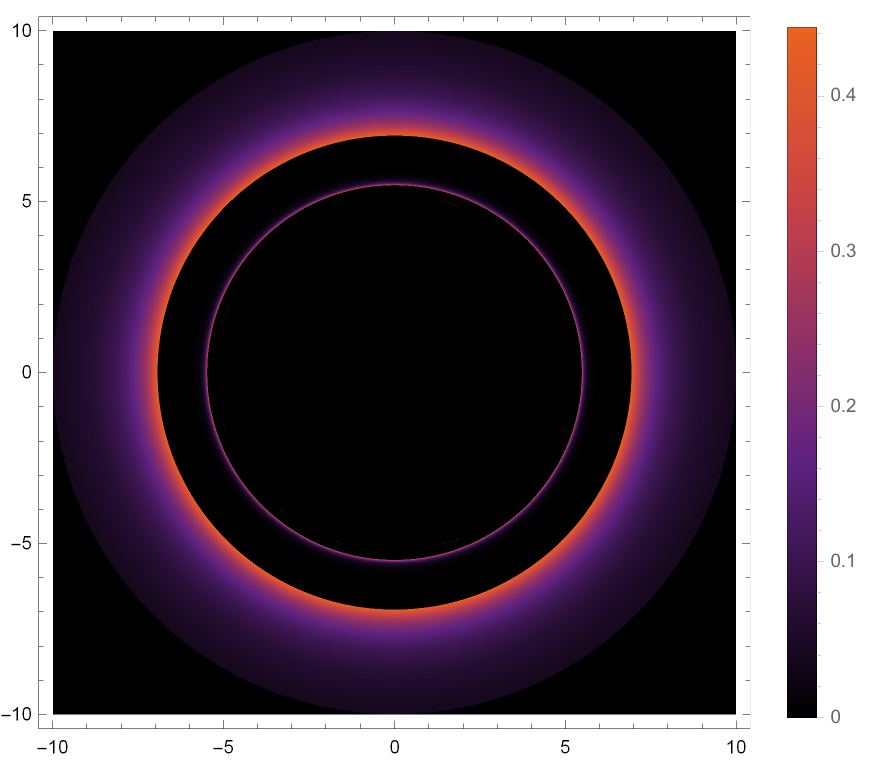}
 \caption{Schwarzschild}
 \end{subfigure}
	\begin{subfigure}{0.3\textwidth}
		\includegraphics[height=3cm, keepaspectratio]{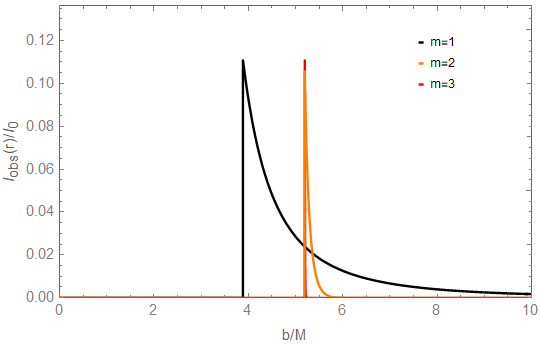}
		\caption{Profile-II}
	\end{subfigure}
	\begin{subfigure}{0.22\textwidth}
		\includegraphics[height=3cm]{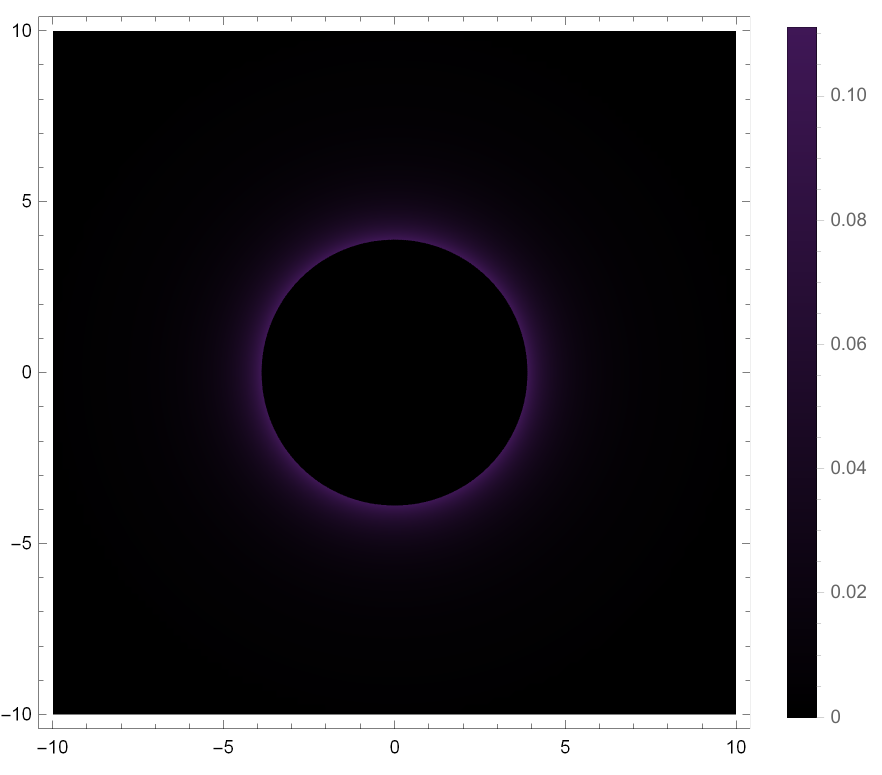}
		\caption{$m=1$}
	\end{subfigure}
	\begin{subfigure}{0.22\textwidth}
		\includegraphics[height=3cm,keepaspectratio]{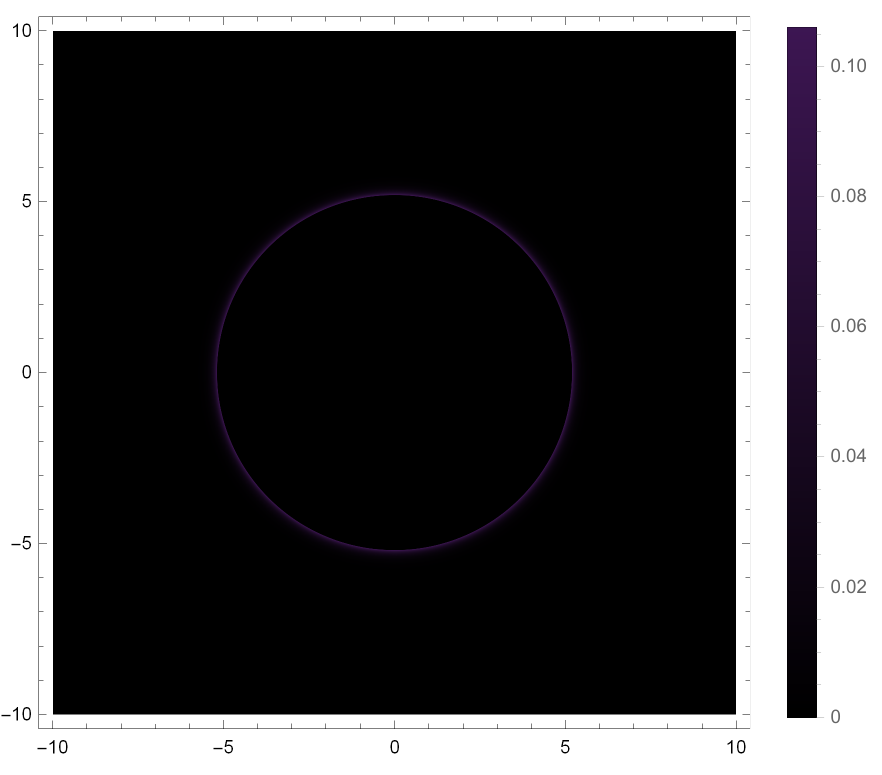}
		\caption{$m=2$}
	\end{subfigure}
 \begin{subfigure}{0.22\textwidth}
 \includegraphics[height=3cm]{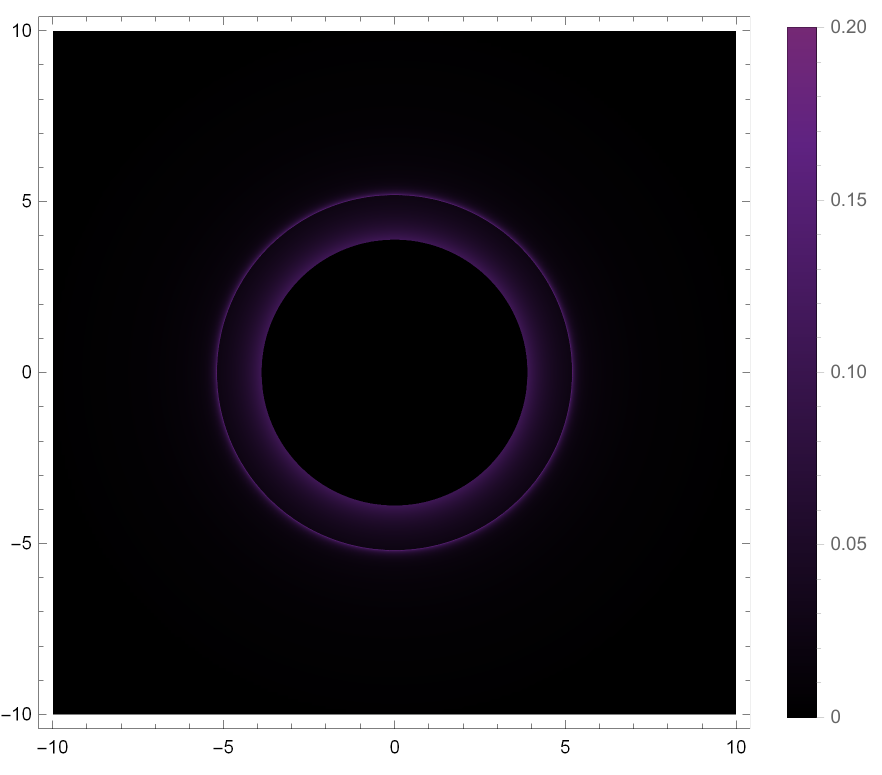}
 \caption{Schwarzschild}
 \end{subfigure}
	\begin{subfigure}{0.3\textwidth}
		\includegraphics[height=3cm, keepaspectratio]{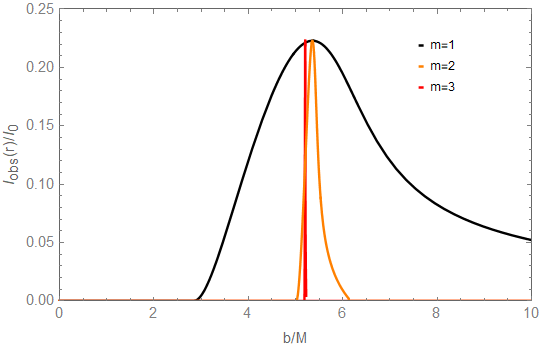}
		\caption{Profile-III}
	\end{subfigure}
	\begin{subfigure}{0.22\textwidth}
		\includegraphics[height=3cm]{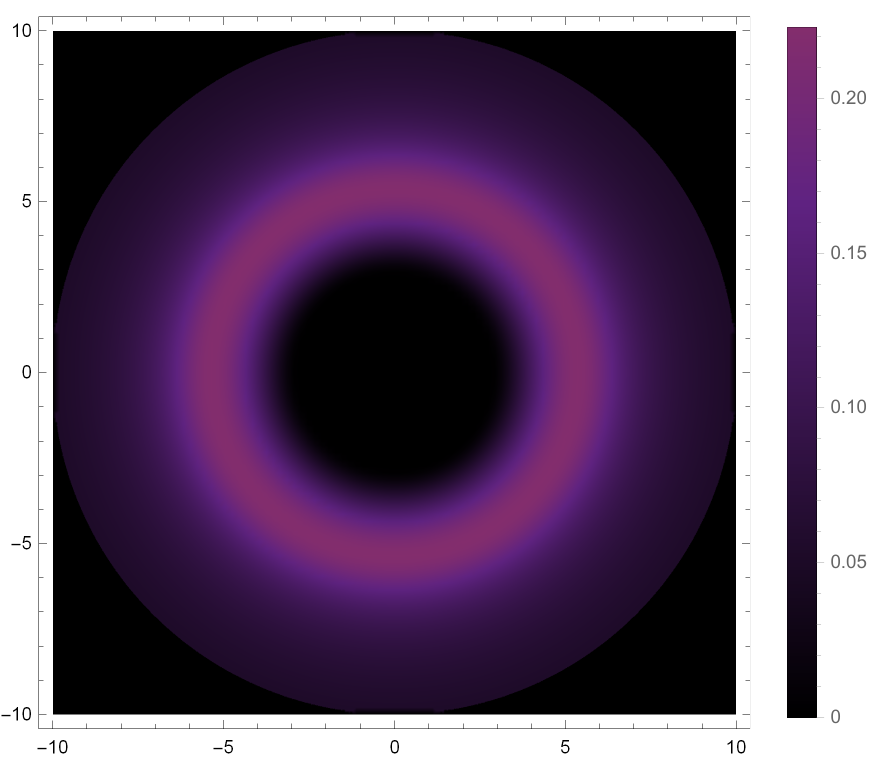}
		\caption{$m=1$}
	\end{subfigure}
	\begin{subfigure}{0.22\textwidth}
		\includegraphics[height=3cm,keepaspectratio]{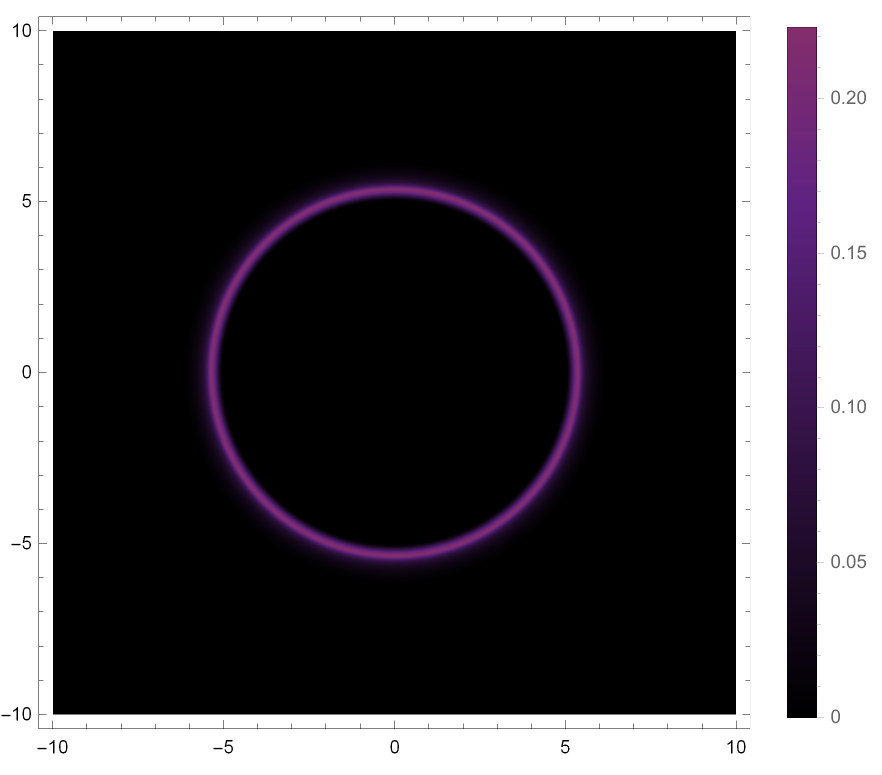}
		\caption{$m=2$}
	\end{subfigure}
	\begin{subfigure}{0.22\textwidth}
		\includegraphics[height=3cm]{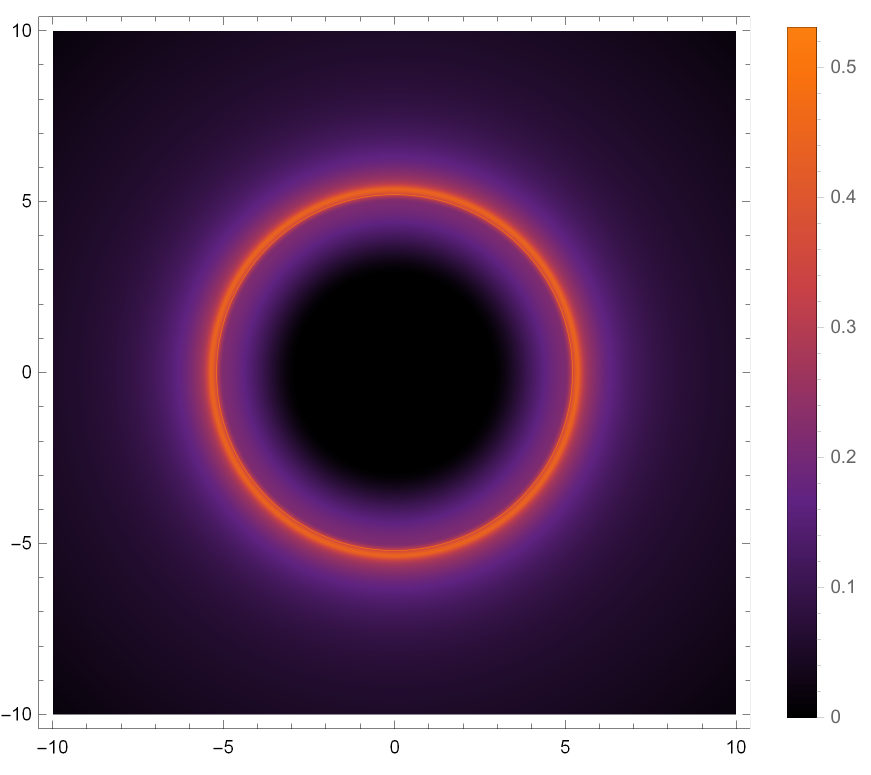}
		\caption{Schwarzschild}
	\end{subfigure}

	\caption{The left column of the figure illustrates the received intensity corresponding to the first three transfer functions ($m=1$, $m=2$, $m=3$) for the Schwarzschild BH. The middle two columns provide the optical appearance in the observer's field of view for the transfer functions ($m=1$, $m=2$). The rightmost column displays the optical appearance of Schwarzschild BH under three types of accretion disks, corresponding to the total received intensity.}
	\label{fig:chuan_guang1}
\end{figure*}

It is worth noting that when using Eq. \eqref{obs} to calculate the received intensity, since the local received intensity is the superposition of contributions from all intersection points with the accretion disk, extreme local brightness may occur. However, here we only consider that the detector measures the average brightness (which is proportional to the measured flux), so there will be no extreme local brightness \cite{Gralla:2019xty}. This also means that when $m>3$, the transfer function corresponds to an extremely small impact parameter, and the flux observed by the detector is very small, making the contribution to the overall observed intensity negligible. To more intuitively visualize the contributions of direct, lensed, and photon ring emission to the total received intensity of the BH, we present the received intensity for different transfer functions ($m=1$, $m=2$, and $m=3$) of the Schwarzschild BH (BH-I and BH-II exhibit similar behavior) in Fig.\ref{fig:chuan_guang1}.

In the left column of Fig.\ref{fig:chuan_guang1}, the black, orange, and red lines represent the received intensity contributed by the transfer functions $m=1$, $m=2$, and $m=3$. Note that the intensities contributed by $m=1$, $m=2$, and $m=3$ represent the direct, the lensed ring, and the photon ring intensity, respectively. Furthermore, we present the brightness distribution of different transfer functions ($m=1$, $m=2$) in the observer's field of view, as well as the optical appearance corresponding to the total received intensity of the Schwarzschild BH. We find that the bright region corresponding to the first transfer function for the three types of accretion disks is brighter and wider in the observer's line of sight compared to the bright region of the second transfer function, indicating that the first transfer function dominates in the optical appearance of the BH. The bright ring corresponding to $m=3$ is extremely thin and nearly invisible, so it is not displayed in this figure.
This also explains why only the first three transfer functions are considered, while higher-order transfer functions are excluded.

\begin{figure*}[htb]
	\centering
	\begin{subfigure}{0.45\textwidth}
		\includegraphics[height=5cm,keepaspectratio]{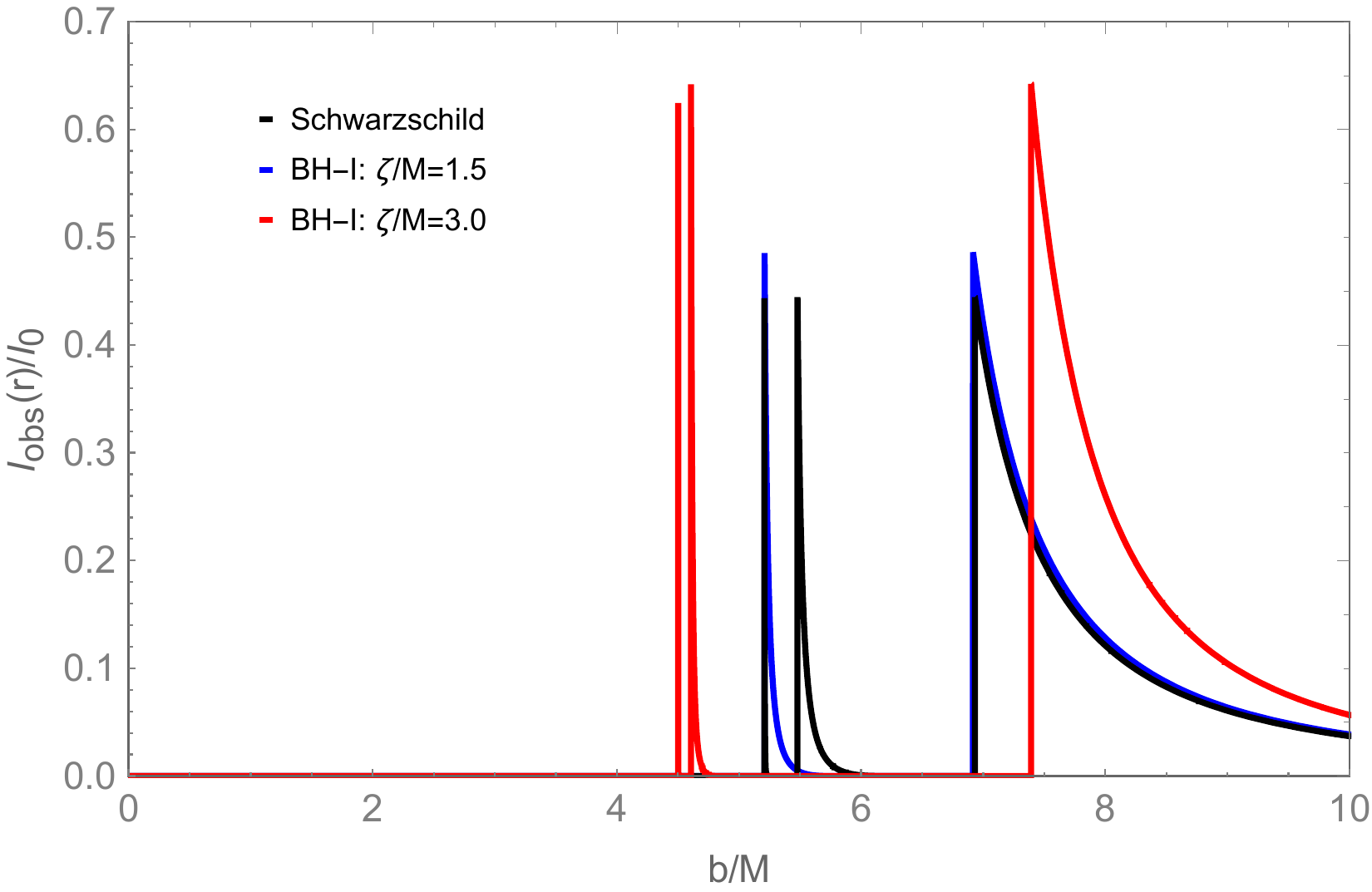}
		\caption{Profile-I}
	\end{subfigure}
	\begin{subfigure}{0.45\textwidth}
		\includegraphics[height=5cm]{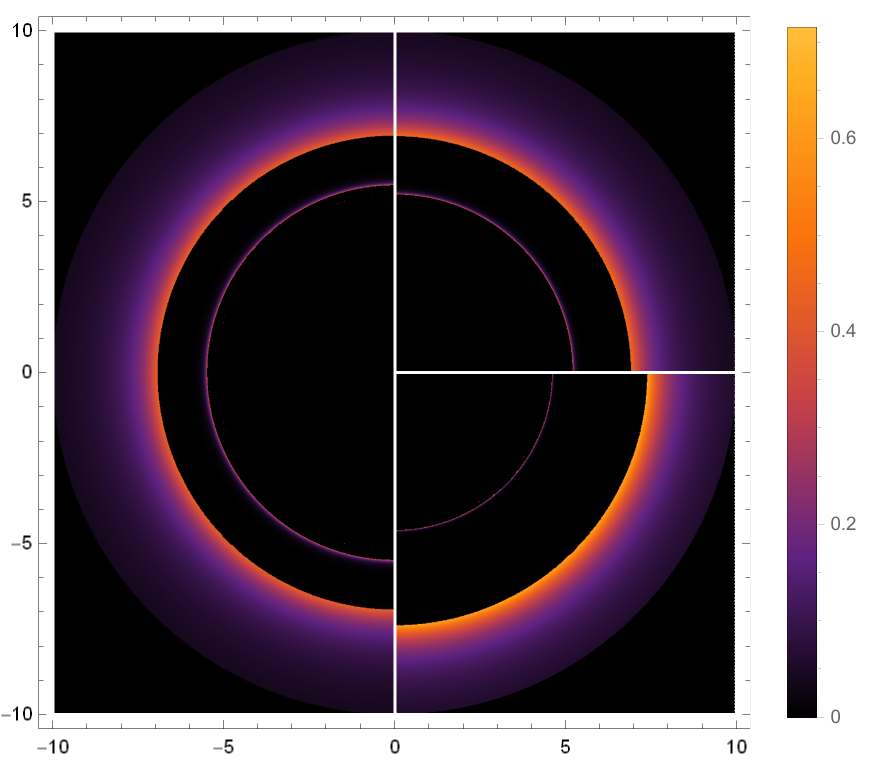}
		\caption{Schwarzschild BH and BH-I}
	\end{subfigure}
	\begin{subfigure}{0.45\textwidth}
		\includegraphics[height=5cm, keepaspectratio]{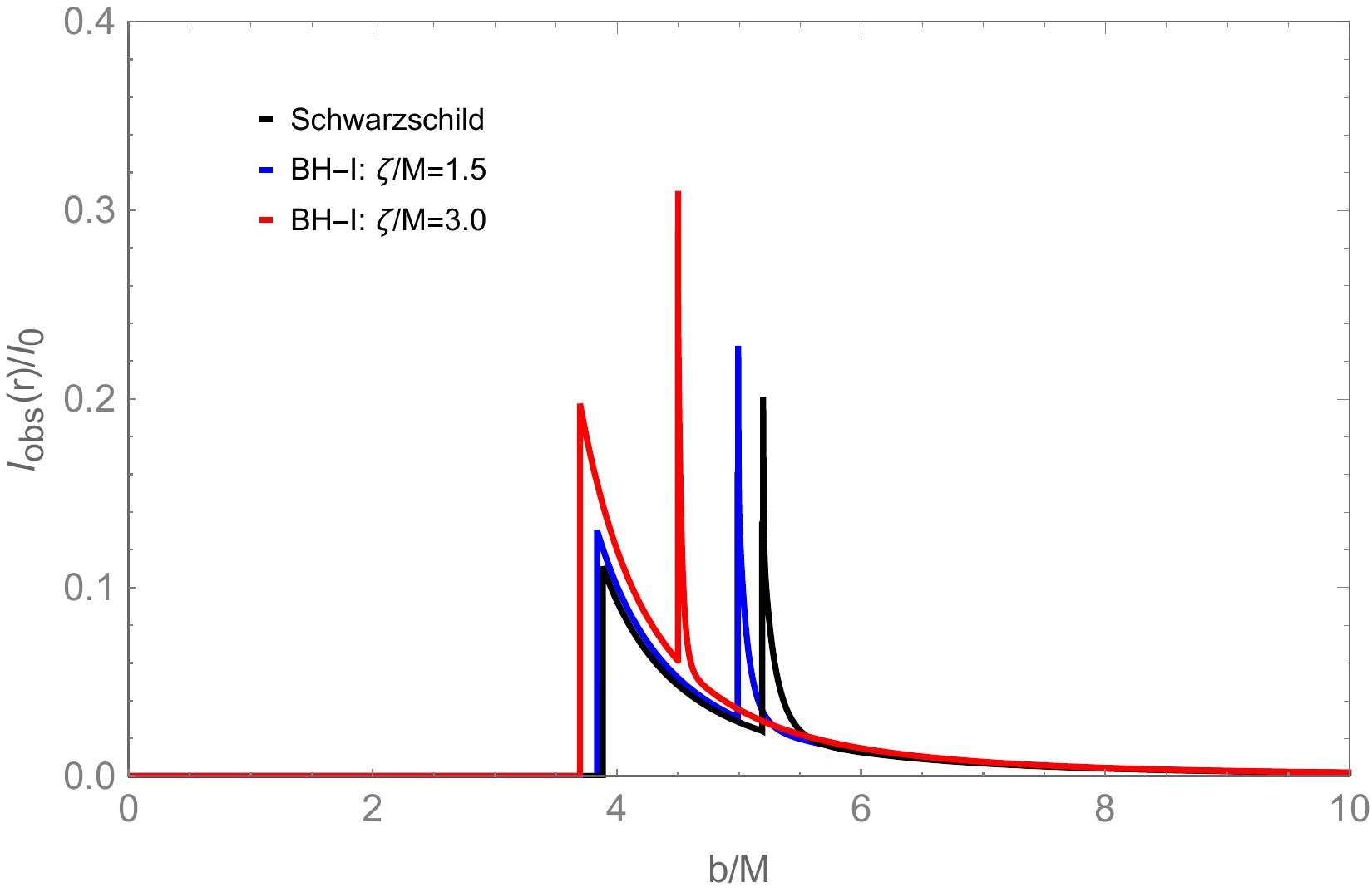}
		\caption{Profile-II}
	\end{subfigure}
	\begin{subfigure}{0.45\textwidth}
		\includegraphics[height=5cm]{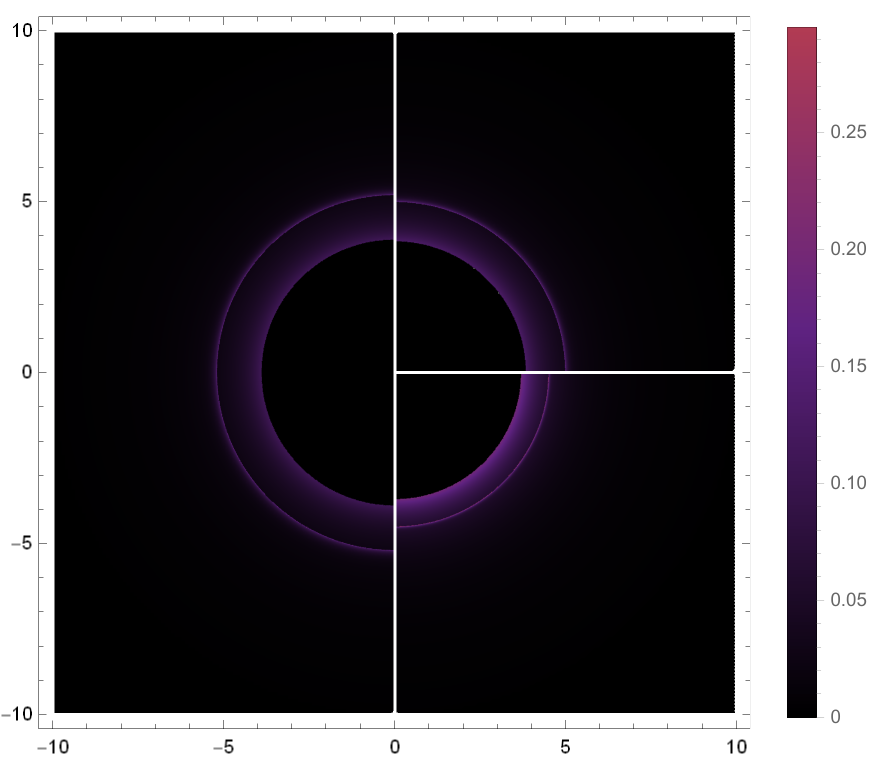}
		\caption{Schwarzschild BH and BH-I}
	\end{subfigure}
	\begin{subfigure}{0.45\textwidth}
		\includegraphics[height=5cm, keepaspectratio]{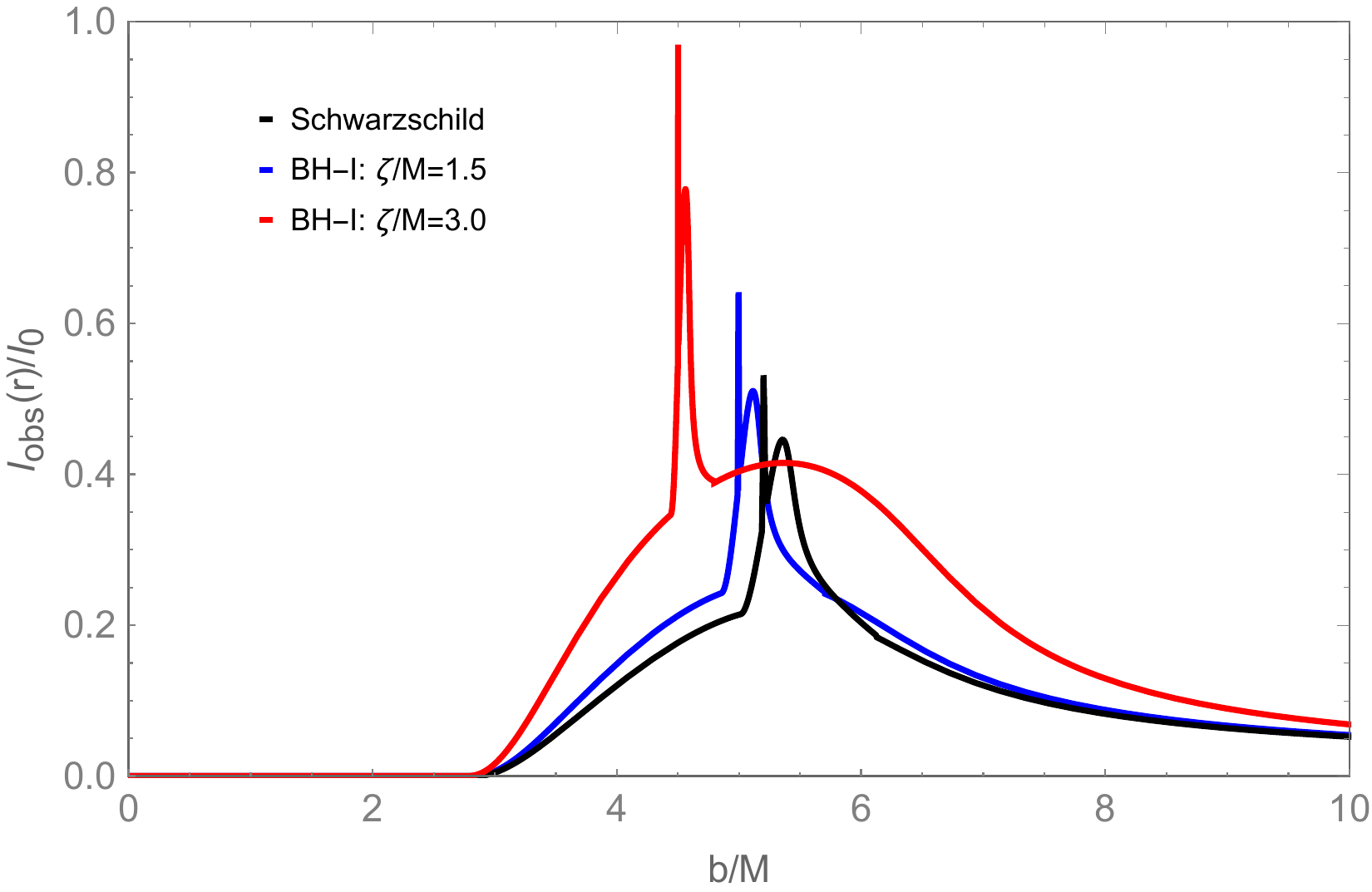}
		\caption{Profile-III}
	\end{subfigure}
	\begin{subfigure}{0.45\textwidth}
		\includegraphics[height=5cm]{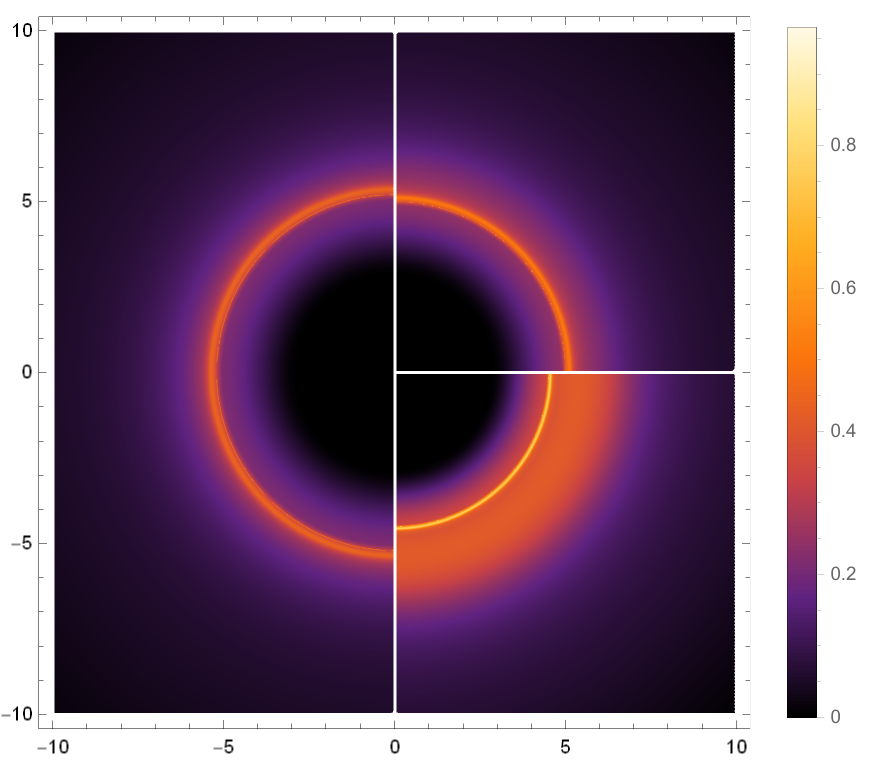}
		\caption{Schwarzschild BH and BH-I}
	\end{subfigure}

	\caption{The figure illustrates the received intensity and optical appearance of three types of accretion disks for BH-I with different values of $\zeta$. The left column shows the received intensity of BH-I, while the right column is composed of images for the Schwarzschild BH (left), $\zeta=1.5$ (top right), and $\zeta=3.0$ (bottom right).}
	\label{fig:guang1}
\end{figure*}

\begin{figure*}[htb]
	\centering
	\begin{subfigure}{0.45\textwidth}
		\includegraphics[height=5cm, keepaspectratio]{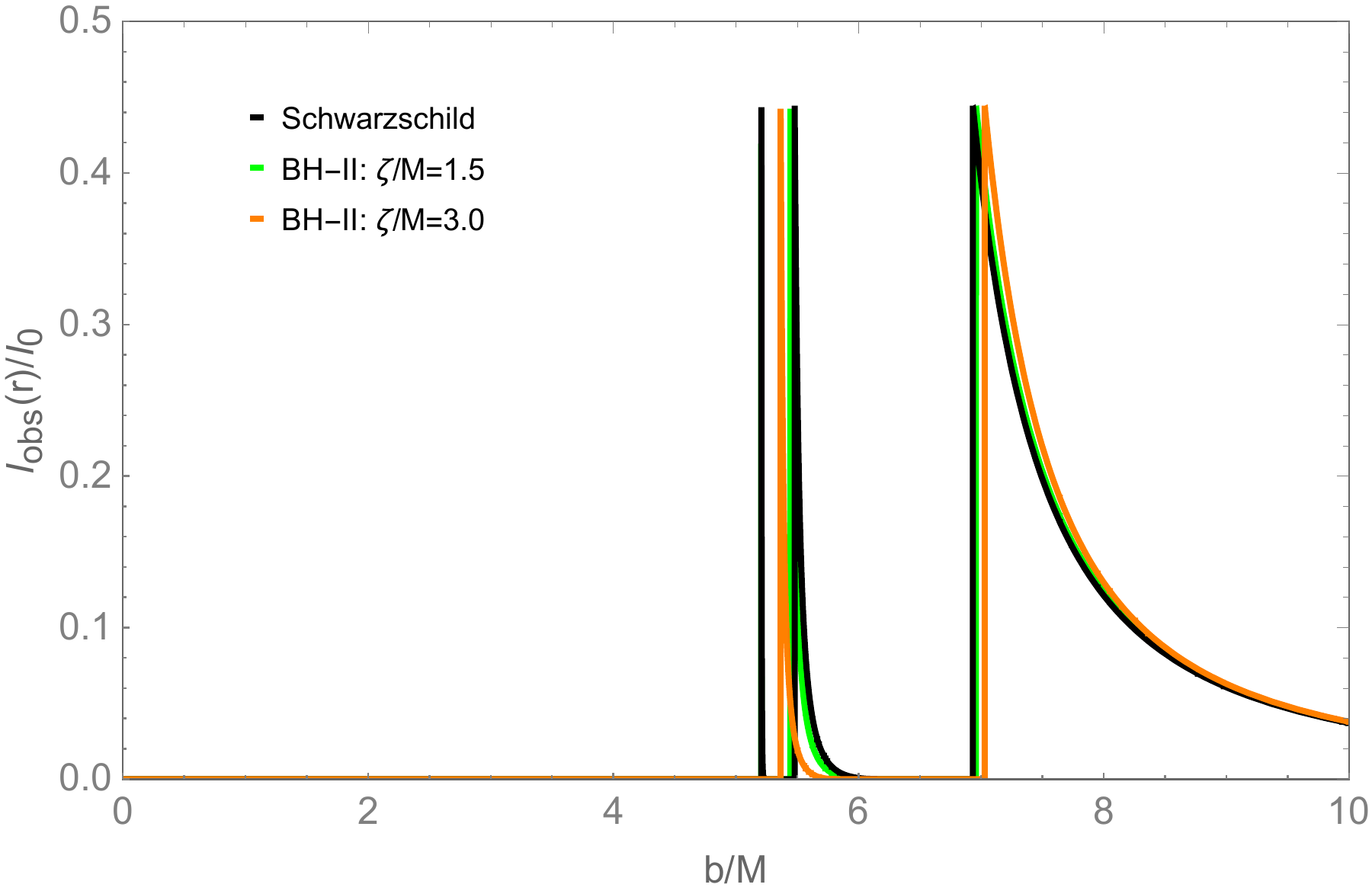}
		\caption{Profile-I}
	\end{subfigure}
	\begin{subfigure}{0.45\textwidth}
		\includegraphics[height=5cm]{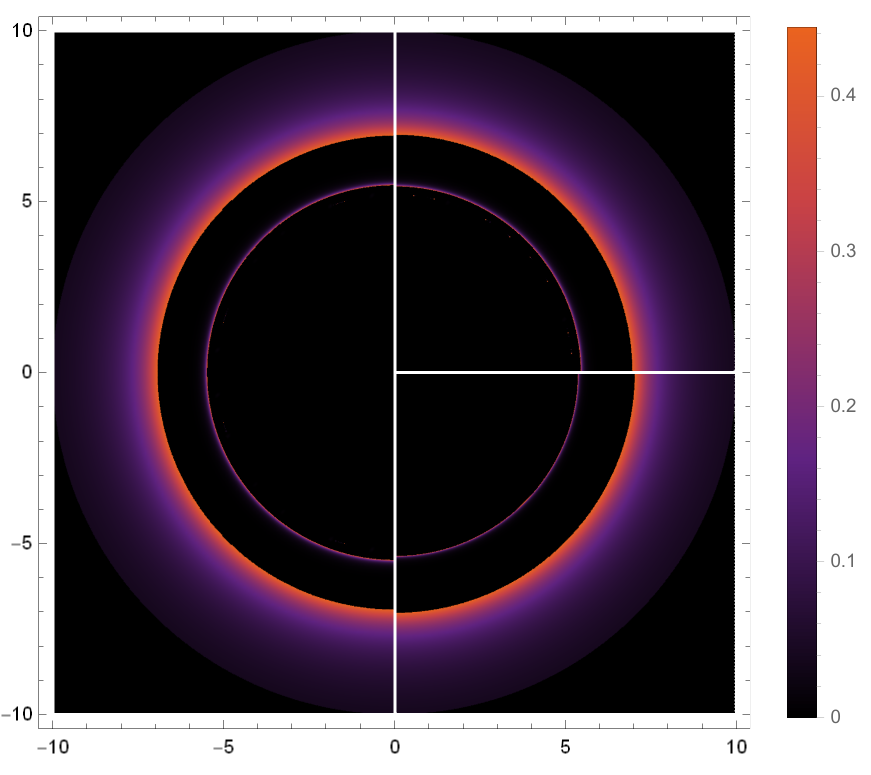}
		\caption{Schwarzschild BH and BH-II}
	\end{subfigure}
	\begin{subfigure}{0.45\textwidth}
		\includegraphics[height=5cm, keepaspectratio]{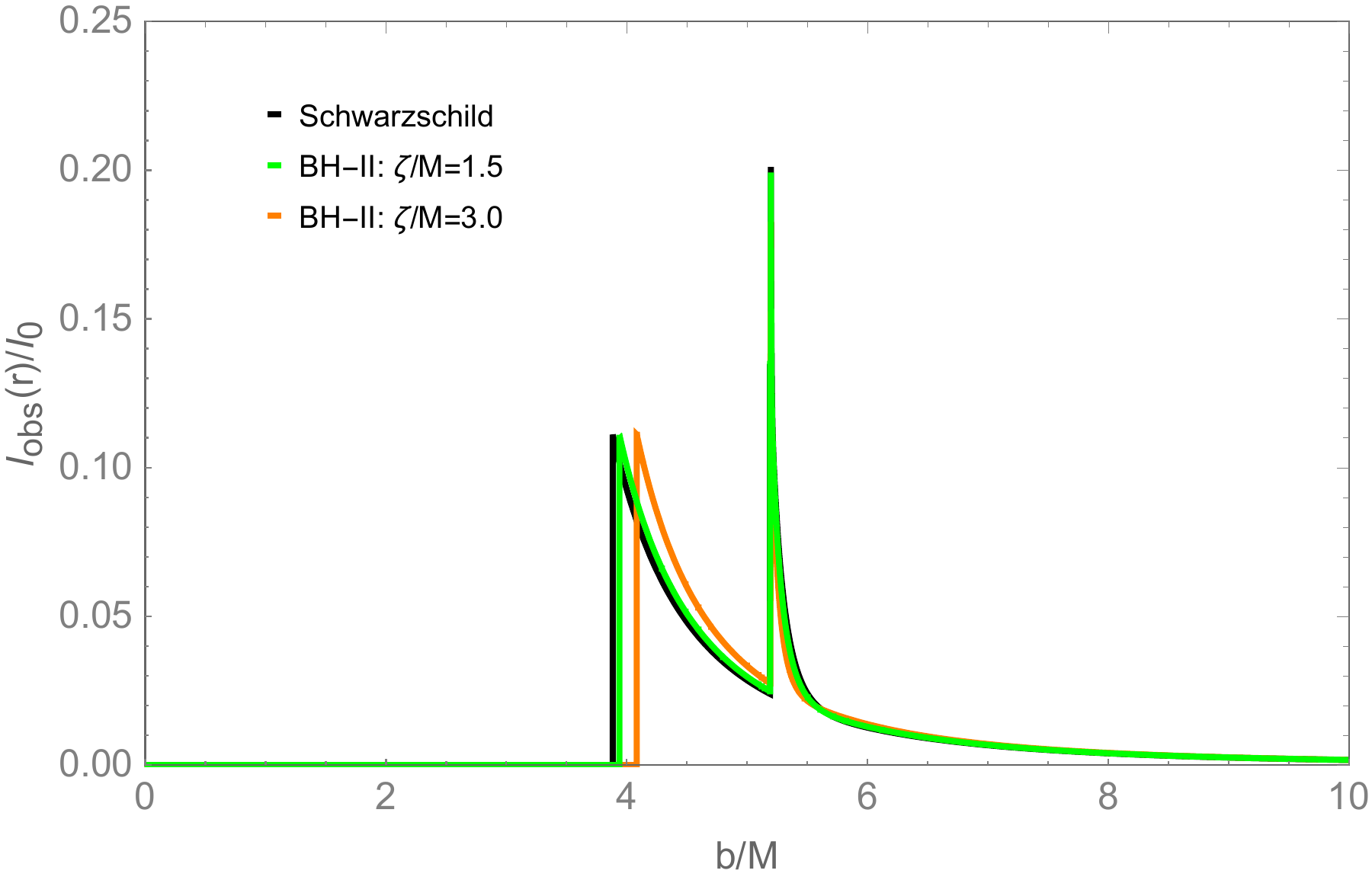}
		\caption{Profile-II}
	\end{subfigure}
	\begin{subfigure}{0.45\textwidth}
		\includegraphics[height=5cm]{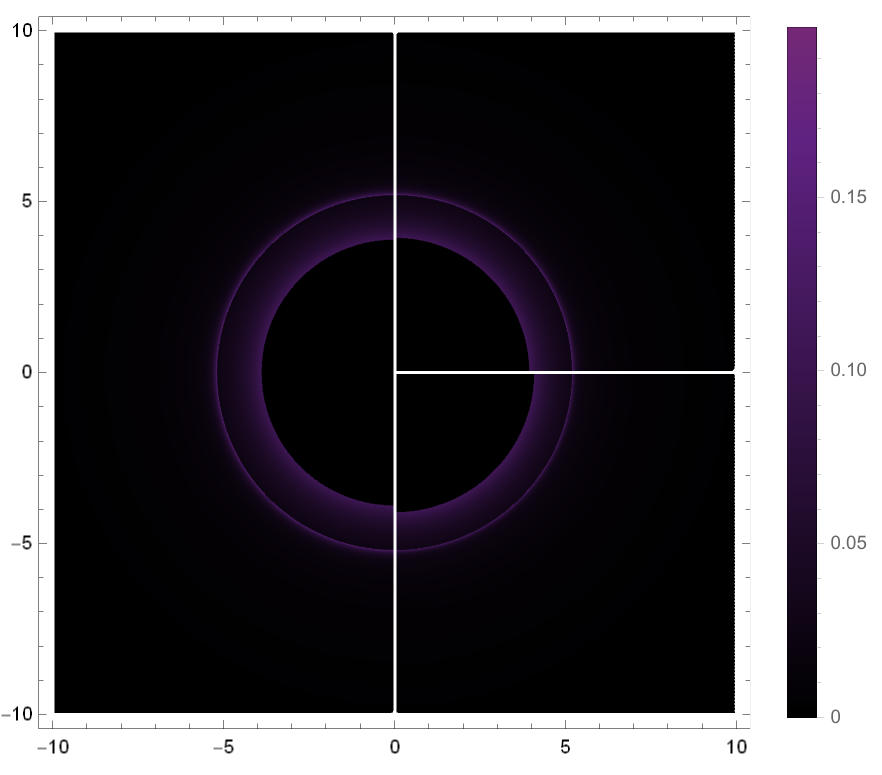}
		\caption{Schwarzschild BH and BH-II}
	\end{subfigure}
	\begin{subfigure}{0.45\textwidth}
		\includegraphics[height=5cm, keepaspectratio]{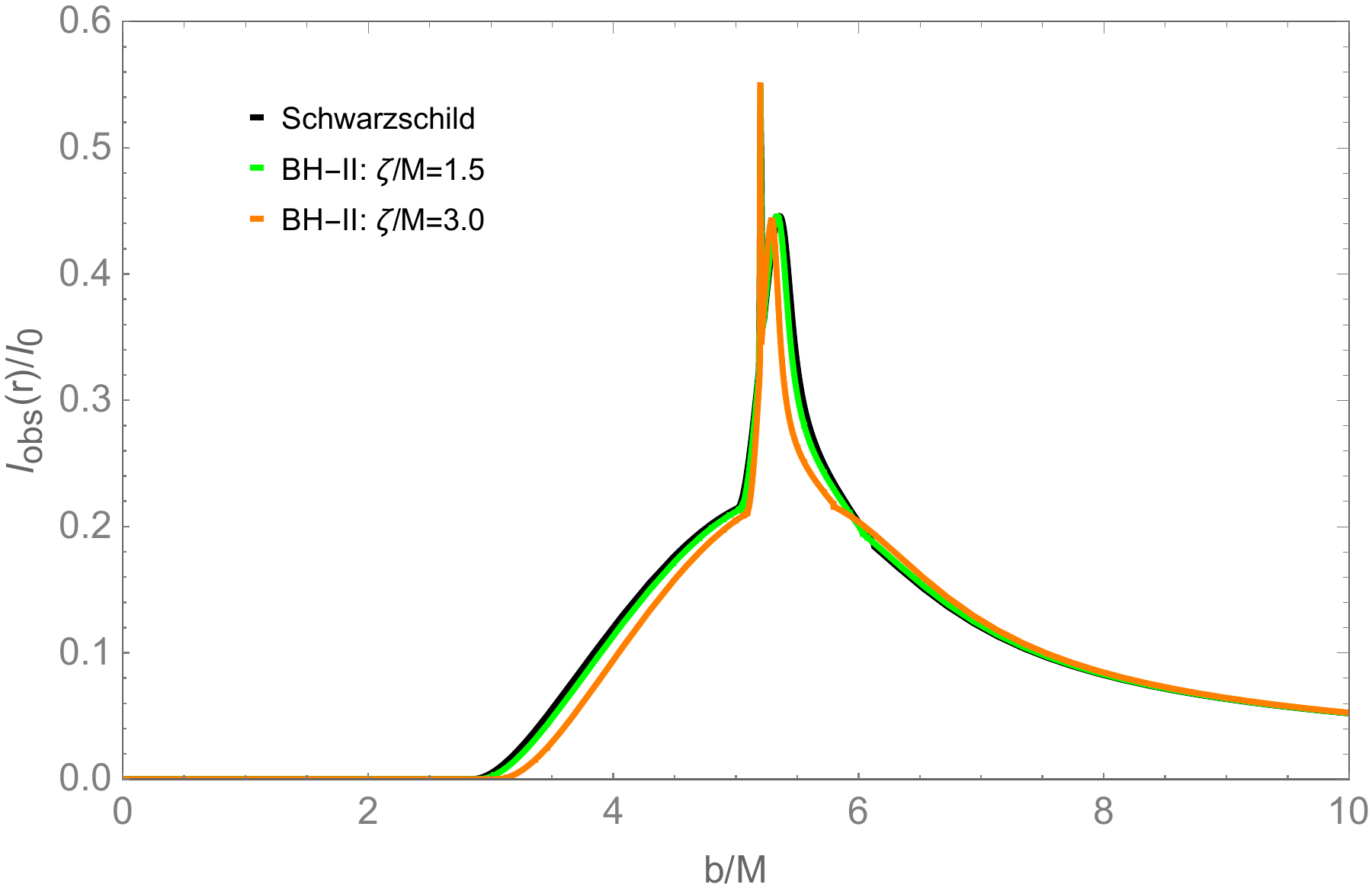}
		\caption{Profile-III}
	\end{subfigure}
	\begin{subfigure}{0.45\textwidth}
		\includegraphics[height=5cm]{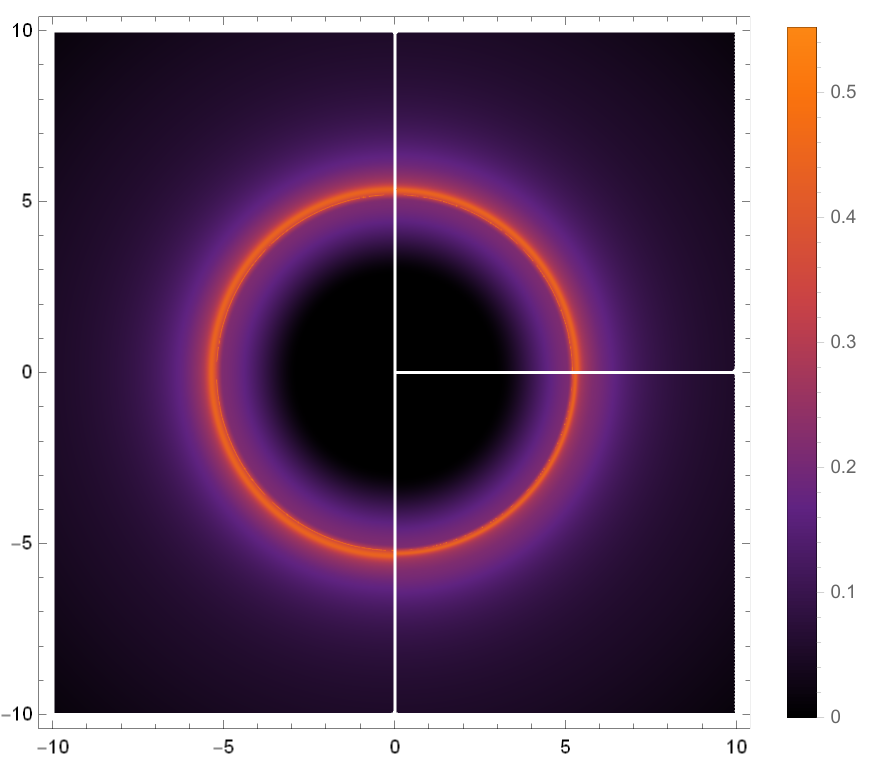}
		\caption{Schwarzschild BH and BH-II}
	\end{subfigure}

	\caption{The figure illustrates the received intensity and optical appearance of three types of accretion disks for BH-II with different values of $\zeta$. The left column shows the received intensity of BH-II, while the right column is composed of images for the Schwarzschild BH (left), $\zeta=1.5$ (top right), and $\zeta=3.0$ (bottom right).}
	\label{fig:guang2}
\end{figure*}

Then, we calculate the total received intensity and represent it in a two-dimensional plane to obtain the optical appearance of two quantum-corrected BHs in Figs. \ref{fig:guang1} and \ref{fig:guang2}. In these plots, the left column shows the received intensity for different types of accretion disks of BH-I and BH-II, while the right column consists of images for the Schwarzschild solution (left), $\zeta=1.5$ (top right), and $\zeta=3.0$ (bottom right). It can be observed that the optical appearance of both quantum-corrected BHs is similar to that of Schwarzschild BH. The introduction of the quantum parameter $\zeta$ only affects the brightness and width of the bright rings.

Focusing on BH-I, the brightness of the rings from all three accretion disk models increases as $\zeta$ increases. Specifically, for the Profile-I, as $\zeta$ increases, the innermost bright ring gradually moves closer to the BH, and its width narrows, while the outer bright ring moves farther away, causing the shadow region between the two bright rings to expand. Turning to the Profile-II, the entire bright ring moves closer to the BH as $\zeta$ increases, and the distance between the two peak bright rings gradually decreases, resulting as a narrowing luminous region in the observer's perspective. Similarly, for the Profile-III, the overall luminous region does not change significantly with $\zeta$, but the peak bright rings narrow and approach the BH, forming an extremely thin and bright ring.
For BH-II, the presence of $\zeta$ has a negligible effect on the brightness of the ring, only slightly influencing the width between the rings. For Profile-I, as $\zeta$ increases, the distance between the two peak bright rings slightly increases. However, for Profile-II, the opposite occurs, with the distance between the bright rings decreasing as $\zeta$ increases. The increase in $\zeta$ also affects the width of the bright rings in Profile-III, making them narrower compared to the Schwarzschild case. Therefore, we can conclude that the impact of $\zeta $ on BH-I is more significant than on BH-II. These optical appearances can help distinguish between the two quantum-corrected BHs.

\section{Rings and images of quantum-corrected BHs with static thin spherical accretions}\label{section4}

The static accretion disk discussed in the previous section is a simplified model. However, in the real universe, the accreting material around a BH is distributed in all directions. Therefore, in this section, we will use a simplified spherical accretion model to simulate the accreting material around the BH and explore its optical appearance in this scenario.

For a BH surrounded by spherical accretion, the radiative intensity observed by a distant observer can be obtained by integrating the radiative specific intensity along the photon path $\gamma$ \cite{Bambi:2012tg,Bambi:2013nla}
\begin{equation}
	I(\nu_o)=\int_{\gamma} g^3 j_e(\nu_e)dl_{\rm prop},\label{eq-Intensity}
\end{equation}
where, $g$ is still the redshift factor, $j_e(\nu_e)$ is the emissivity of the source in the rest frame per unit volume, usually taken as $j_e(\nu_e)\propto \delta(\nu_e-\nu_\star)/r^2$($\nu_\star$ is the rest-frame frequency). And $dl_{\rm prop}$ is the infinitesimal proper length of the photon path in the rest frame of the emitter. The total observed intensity by the observer can be obtained by integrating Eq. \eqref{eq-Intensity}, i.e.
\begin{equation}
	I_{\rm obs}=\int_{\nu_e}\int_\gamma g^4 j_e(\nu_e)dl_{\rm prop} d\nu_e=\int_\gamma \frac{g^4}{r^2} dl_{\rm prop}.\label{eq-Intensity2}
\end{equation}

\begin{figure*}[htb]
	\centering
	\begin{subfigure}{0.45\textwidth}
		\includegraphics[height=5cm, keepaspectratio]{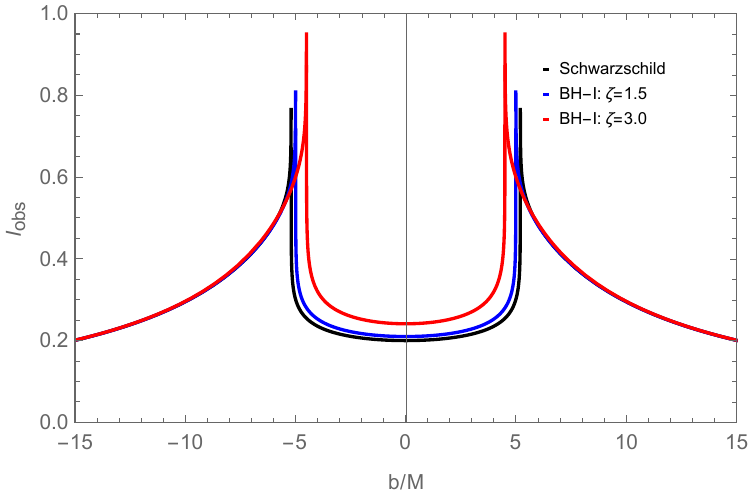}
		\caption{$I_{\rm obs}$}
	\end{subfigure}
	\begin{subfigure}{0.45\textwidth}
		\includegraphics[height=5cm]{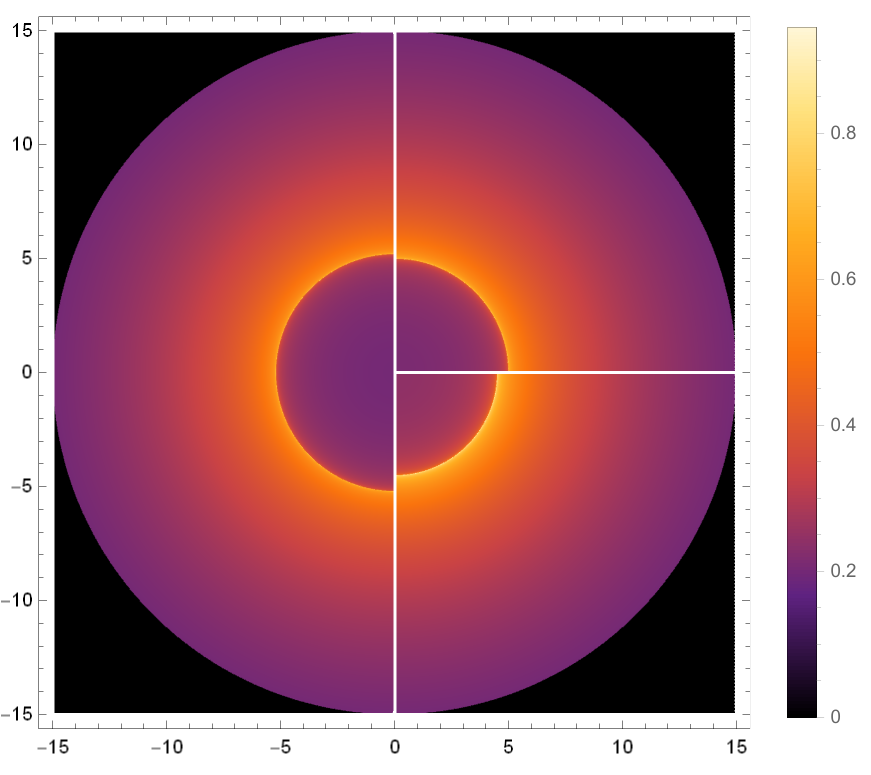}
		\caption{Schwarzschild BH and BH-I}
	\end{subfigure}
	\begin{subfigure}{0.45\textwidth}
		\includegraphics[height=5cm, keepaspectratio]{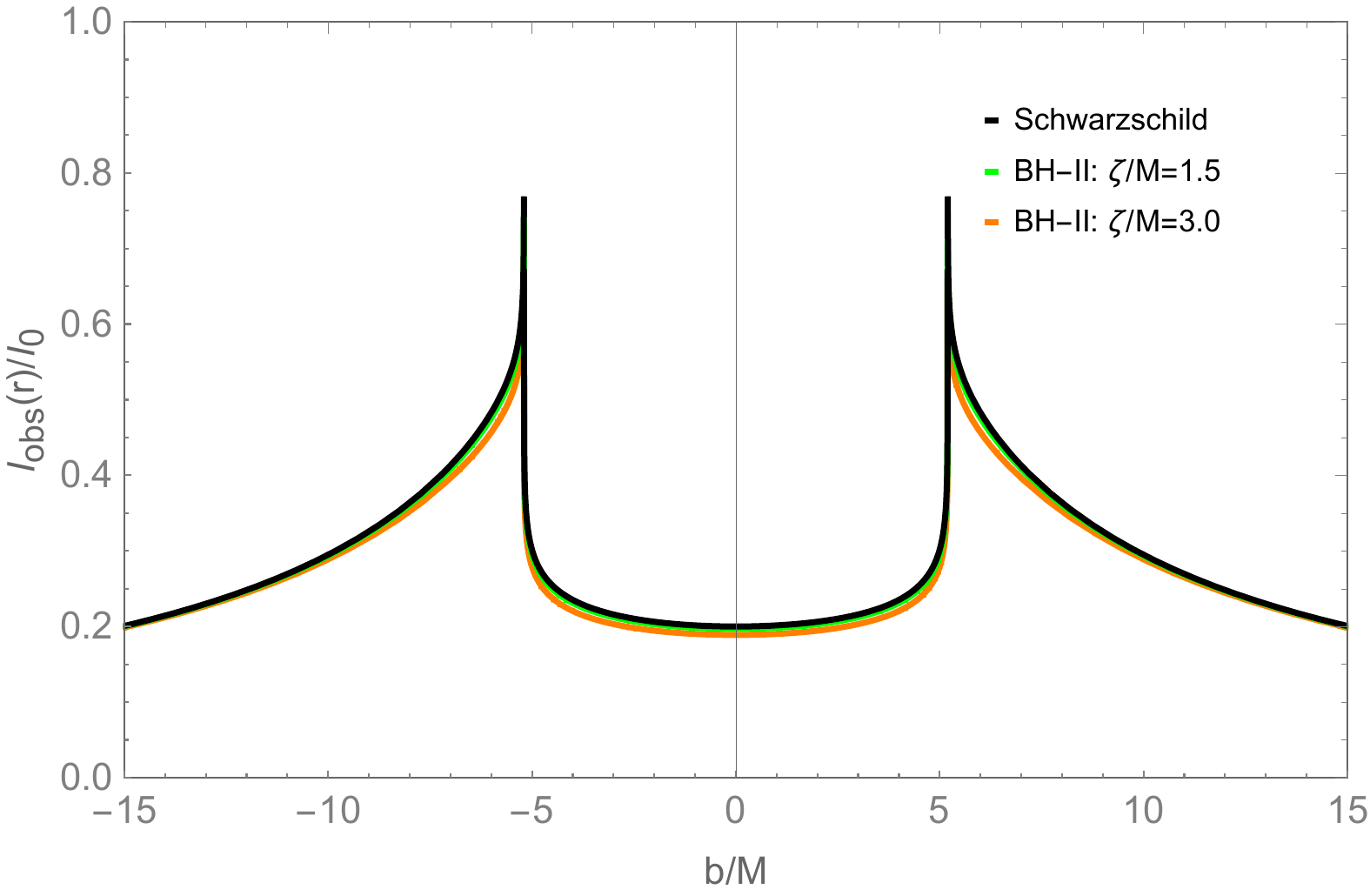}
		\caption{$I_{\rm obs}$}
	\end{subfigure}
	\begin{subfigure}{0.45\textwidth}
		\includegraphics[height=5cm]{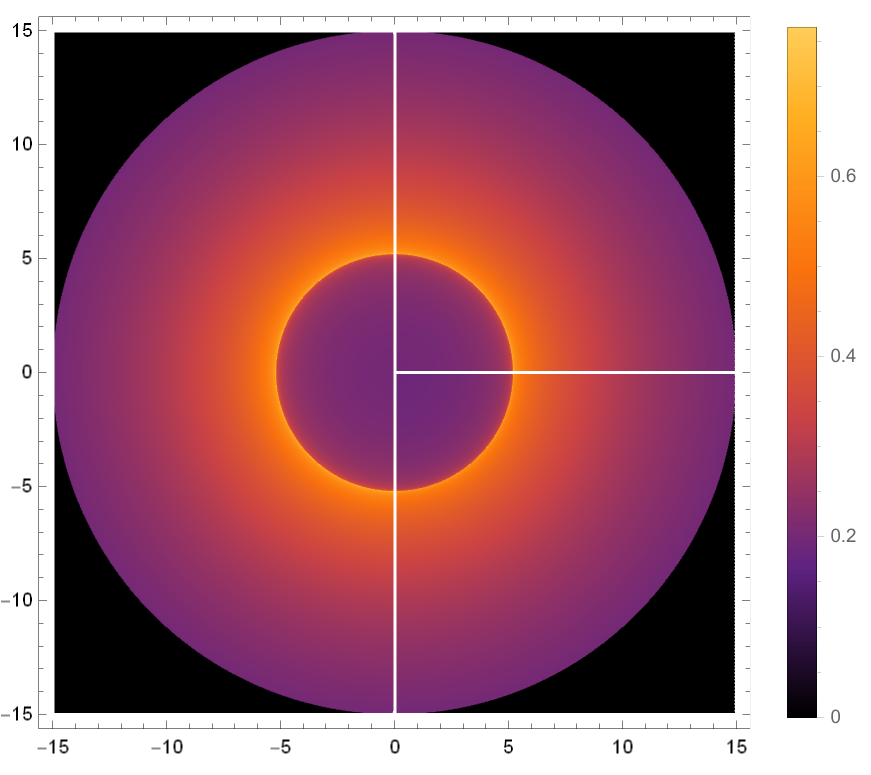}
		\caption{Schwarzschild BH and BH-II}
	\end{subfigure}%

	\caption{The figure shows the total received intensity and optical appearance of BH-I and BH-II for static thin spherical accretions. The left column displays the received intensities for both BH-I and BH-II, while the right column is composed of images for the Schwarzschild BH (left), $\zeta=1.5$ (top right), and $\zeta=3.0$ (bottom right).}\label{fig:static}
\end{figure*}

For a static spherical accretion, the $dl_{\rm prop}$ can be expressed as
\begin{equation}
	dl_{\rm prop}=\sqrt{h_{\rm ij}d x^i dx^j}=\sqrt{\frac{1}{g(r)}+h(r)\left(\frac{d\phi}{dr}\right)^2}dr,
\end{equation}
here, $h_{\rm ij}$ represents the induced metric of the static spherically symmetric metric. Combining with Eq. \eqref{ef}, we obtain the final total observed intensity as:
\begin{equation}
	I_{\rm obs}=\int_\gamma \frac{g^4}{r^2} \sqrt{\frac{1}{g(r)}+\frac{b^2}{r^2-b^2 f(r)} \frac{f(r)}{g(r)}}dr.\label{iob_static}
\end{equation}
In Eq. \eqref{iob_static}, by specifying the impact parameter $b$, we can integrate along different photon paths to obtain the total observed intensity and the optical appearance in Fig. \ref{fig:static}.

The left column of the figure illustrates the variation of the total observed intensity with the impact parameter, where a peak always appears at $b = b_{\text{ph}}$, corresponding to photons reaching the observer after multiple orbits. The right columns show the 2D optical appearances of the corresponding BHs. The dim central region represents the BH shadow, while the surrounding luminous area corresponds to the spherical accretion.

For BH-I, as $\zeta$ increases, the overall observed intensity curve shifts upward and gradually contracts, resulting in the bright ring region becoming brighter and the dark region slightly shrinking in the optical appearance. In contrast, for BH-II, the influence of $\zeta$ is not significant, making it almost indistinguishable from Schwarzschild BH. This is consistent with the previous discussion.

\section{Conclusion}\label{section5}

In this paper, we mainly study the impact of the quantum parameter $ \zeta $ on the optical appearance of BHs under different accretion models. After discussing the geodesic motion of a massive particle around the two quantum-corrected BHs (BH-I and BH-II), we calculated the variation of $r_{\rm h}$, $r_{\text{ph}}$, $b_{\text{ph}}$, and $r_{\text{isco}}$ with the quantum parameter. It was found that only $b_{\text{ph}}$ and $r_{\text{isco}}$ for BH-I change with the quantum parameter. Then, using the angular shadow diameters of M87* and Sgr A*, we provided the constraint range for the quantum parameter $ \zeta $ of BH-I: $ 0 \leq \frac{\zeta}{M} \leq 4.74 $ for M87* and $ 0 \leq \frac{\zeta}{M} \leq 3.52 $ for Sgr A*. For BH-II, it cannot be strictly constrained by the known observational data.

Within the above constraint range, we further examined the photon trajectories near the quantum-corrected BHs for different values of $ \zeta $. We found that the introduction of $ \zeta $ decreases the range of impact parameters corresponding to the photon ring and lensed ring, with the effect being more significant for BH-I than for BH-II. Subsequently, we studied the optical appearance of BH-I and BH-II under three types of geometrically thin accretion disks. In this process, we presented the contributions of different transfer functions ($m=1$, $m=2$, and $m=3$) to the received intensity. We have also intuitively illustrated the luminosity of direct and lensed ring corresponding to these transfer functions within the observer's field of view (see Fig. \ref{fig:chuan_guang1}). From these figures, we observed that the intensity of direct emission ($m=1$) makes the primary contribution to the BH image in all three accretion disk models. We then displayed the optical appearances of BH-I and BH-II for different values of $ \zeta $ under the three types of accretion disks in Fig. \ref{fig:guang1} and \ref{fig:guang2}. It is evident that $ \zeta $ causes changes in the brightness and width of the bright ring, especially for BH-I. The optical appearance of BH-I becomes brighter as $ \zeta $ increases, and the photon ring moves closer to the BH. However, $\zeta$ only affects the distance between the bright rings of BH-II, with its effect on the brightness being almost negligible.

The static thin spherical accretions around BH-I and BH-II are also considered. We found that $ \zeta $ still influences the brightness and width of the bright ring for BH-I. As $ \zeta $ increases, the optical appearance of BH-I becomes brighter, and the central dark region becomes smaller. For BH-II, the effect of $ \zeta $ on the bright ring is minimal.

In conclusion, our results indicate that the optical appearance of BH-II is nearly indistinguishable from that of the Schwarzschild BH, with only a slight difference emerging when $ \zeta $ is very large. This suggests that there may be some form of degeneracy between BH-II and the Schwarzschild BH. However, the optical appearance of BH-I shows a significant difference from that of the Schwarzschild BH, providing a potential method for distinguishing between these models from an observational perspective in the future.

\begin{acknowledgments}
This work is supported in part by NSFC Grant No. 12165005.
\end{acknowledgments}




\begin{thebibliography}{84}

\bibitem{LIGO:2017dbh}
 B.P. Abbott {et~al.} (LIGO Scientific Collaboration and Virgo Collaboration),
 {Observation of Gravitational Waves from a Binary Black Hole Merger},
 in \emph{{Centennial of General Relativity}: {A Celebration}},
 edited by C.A.Z. Vasconcellos
 (World Scientific, Singapore, 2017).
 {\url{https://doi.org/10.1142/9789814699662_0011}}

\bibitem{EventHorizonTelescope:2019dse}
 K.~Akiyama {et~al.} (Event Horizon Telescope Collaboration),
 {First M87 Event Horizon Telescope results. I. The shadow of the supermassive
 black hole}.
 Astrophys. J. Lett. \textbf{875}, L1 (2019).
 {\url{https://doi.org/10.3847/2041-8213/ab0ec7}}.
 {\href{https://arxiv.org/abs/1906.11238}{{arXiv:1906.11238}}}

\bibitem{Penrose:1964wq}
 R.~Penrose,
 {Gravitational collapse and space-time singularities}.
 Phys. Rev. Lett. \textbf{14}, 57 (1965).
 {\url{https://doi.org/10.1103/PhysRevLett.14.57}}

\bibitem{Hawking:1970zqf}
 S.W. Hawking, R.~Penrose,
 {The singularities of gravitational collapse and cosmology}.
 Proc. Roy. Soc. Lond. A \textbf{314}, 529 (1970).
 {\url{https://doi.org/10.1098/rspa.1970.0021}}

\bibitem{Rovelli:2011eq}
 C.~Rovelli,
 {Zakopane lectures on loop gravity}.
 PoS \textbf{QGQGS2011}, 003 (2011).
 {\url{https://doi.org/10.22323/1.140.0003}}.
 {\href{https://arxiv.org/abs/1102.3660}{{arXiv:1102.3660}}}

\bibitem{Bojowald:2008zzb}
 M.~Bojowald,
 {Loop quantum cosmology}.
 Living Rev. Rel. \textbf{11}, 4 (2008).
 {\url{https://doi.org/10.12942/lrr-2008-4}}

\bibitem{Rovelli:1997yv}
 C.~Rovelli,
 {Loop quantum gravity}.
 Living Rev. Rel. \textbf{1}, 1 (1998).
 {\url{https://doi.org/10.12942/lrr-1998-1}}.
 {\href{https://arxiv.org/abs/gr-qc/9710008}{{arXiv:gr-qc/9710008}}}

\bibitem{Ashtekar:2005qt}
 A.~Ashtekar, M.~Bojowald,
 {Quantum geometry and the Schwarzschild singularity}.
 Class. Quant. Grav. \textbf{23}, 391 (2006).
 {\url{https://doi.org/10.1088/0264-9381/23/2/008}}.
 {\href{https://arxiv.org/abs/gr-qc/0509075}{{arXiv:gr-qc/0509075}}}

\bibitem{Ashtekar:2004eh}
 A.~Ashtekar, J.~Lewandowski,
 {Background independent quantum gravity: A status report}.
 Class. Quant. Grav. \textbf{21}, R53 (2004).
 {\url{https://doi.org/10.1088/0264-9381/21/15/R01}}.
 {\href{https://arxiv.org/abs/gr-qc/0404018}{{arXiv:gr-qc/0404018}}}

\bibitem{Ashtekar:2013hs}
 A.~Ashtekar,
 {Introduction to loop quantum gravity and cosmology}.
 Lect. Notes Phys. \textbf{863}, 31 (2013).
 {\url{https://doi.org/10.1007/978-3-642-33036-0_2}}.
 {\href{https://arxiv.org/abs/1201.4598}{{arXiv:1201.4598}}}

\bibitem{Thiemann:2001gmi}
 T.~Thiemann,
 {Modern canonical quantum general relativity}.
 {\href{https://arxiv.org/abs/gr-qc/0110034}{{arXiv:gr-qc/0110034}}}

\bibitem{Han:2005km}
 M.~Han, W.~Huang, Y.~Ma,
 {Fundamental structure of loop quantum gravity}.
 Int. J. Mod. Phys. D \textbf{16}, 1397 (2007).
 {\url{https://doi.org/10.1142/S0218271807010894}}.
 {\href{https://arxiv.org/abs/gr-qc/0509064}{{arXiv:gr-qc/0509064}}}

\bibitem{Modesto:2008im}
 L.~Modesto,
 {Semiclassical loop quantum black hole}.
 Int. J. Theor. Phys. \textbf{49}, 1649 (2010).
 {\url{https://doi.org/10.1007/s10773-010-0346-x}}.
 {\href{https://arxiv.org/abs/0811.2196}{{arXiv:0811.2196}}}

\bibitem{Perez:2017cmj}
 A.~Perez,
 {Black holes in loop quantum gravity}.
 Rept. Prog. Phys. \textbf{80}, 126901 (2017).
 {\url{https://doi.org/10.1088/1361-6633/aa7e14}}.
 {\href{https://arxiv.org/abs/1703.09149}{{arXiv:1703.09149}}}

\bibitem{Ashtekar:2018lag}
 A.~Ashtekar, J.~Olmedo, P.~Singh,
 {Quantum transfiguration of Kruskal black holes}.
 Phys. Rev. Lett. \textbf{121}, 241301 (2018).
 {\url{https://doi.org/10.1103/PhysRevLett.121.241301}}.
 {\href{https://arxiv.org/abs/1806.00648}{{arXiv:1806.00648}}}

\bibitem{Bodendorfer:2019cyv}
 N.~Bodendorfer, F.M. Mele, J.~M\"unch,
 {Effective quantum extended spacetime of polymer Schwarzschild black hole}.
 Class. Quant. Grav. \textbf{36}, 195015 (2019).
 {\url{https://doi.org/10.1088/1361-6382/ab3f16}}.
 {\href{https://arxiv.org/abs/1902.04542}{{arXiv:1902.04542}}}

\bibitem{Kelly:2020uwj}
 J.G. Kelly, R.~Santacruz, E.~Wilson-Ewing,
 {Effective loop quantum gravity framework for vacuum spherically symmetric
 spacetimes}.
 Phys. Rev. D \textbf{102}, 106024 (2020).
 {\url{https://doi.org/10.1103/PhysRevD.102.106024}}.
 {\href{https://arxiv.org/abs/2006.09302}{{arXiv:2006.09302}}}

\bibitem{Gan:2020dkb}
 W.C. Gan, N.O. Santos, F.W. Shu, A.~Wang,
 {Properties of the spherically symmetric polymer black holes}.
 Phys. Rev. D \textbf{102}, 124030 (2020).
 {\url{https://doi.org/10.1103/PhysRevD.102.124030}}.
 {\href{https://arxiv.org/abs/2008.09664}{{arXiv:2008.09664}}}

\bibitem{Sartini:2020ycs}
 F.~Sartini, M.~Geiller,
 {Quantum dynamics of the black hole interior in loop quantum cosmology}.
 Phys. Rev. D \textbf{103}, 066014 (2021).
 {\url{https://doi.org/10.1103/PhysRevD.103.066014}}.
 {\href{https://arxiv.org/abs/2010.07056}{{arXiv:2010.07056}}}

\bibitem{Song:2020arr}
 S.~Song, H.~Li, Y.~Ma, C.~Zhang,
 {Entropy of black holes with arbitrary shapes in loop quantum gravity}.
 Sci. China Phys. Mech. Astron. \textbf{64}, 120411 (2021).
 {\url{https://doi.org/10.1007/s11433-021-1770-3}}.
 {\href{https://arxiv.org/abs/2002.08869}{{arXiv:2002.08869}}}

\bibitem{Zhang:2020qxw}
 C.~Zhang, Y.~Ma, S.~Song, X.~Zhang,
 {Loop quantum Schwarzschild interior and black hole remnant}.
 Phys. Rev. D \textbf{102}, 041502 (2020).
 {\url{https://doi.org/10.1103/PhysRevD.102.041502}}.
 {\href{https://arxiv.org/abs/2006.08313}{{arXiv:2006.08313}}}

\bibitem{Zhang:2021wex}
 C.~Zhang, Y.~Ma, S.~Song, X.~Zhang,
 {Loop quantum deparametrized Schwarzschild interior and discrete black hole
 mass}.
 Phys. Rev. D \textbf{105}, 024069 (2022).
 {\url{https://doi.org/10.1103/PhysRevD.105.024069}}.
 {\href{https://arxiv.org/abs/2107.10579}{{arXiv:2107.10579}}}

\bibitem{Lewandowski:2022zce}
 J.~Lewandowski, Y.~Ma, J.~Yang, C.~Zhang,
 {Quantum Oppenheimer-Snyder and Swiss Cheese models}.
 Phys. Rev. Lett. \textbf{130}, 101501 (2023).
 {\url{https://doi.org/10.1103/PhysRevLett.130.101501}}.
 {\href{https://arxiv.org/abs/2210.02253}{{arXiv:2210.02253}}}

\bibitem{Lewandowski:2021bkt}
 J.~Lewandowski, C.~Zhang,
 {Fermion coupling to loop quantum gravity: Canonical formulation}.
 Phys. Rev. D \textbf{105}, 124025 (2022).
 {\url{https://doi.org/10.1103/PhysRevD.105.124025}}.
 {\href{https://arxiv.org/abs/2112.08865}{{arXiv:2112.08865}}}

\bibitem{Zhang:2022vsl}
 C.~Zhang, H.~Liu, M.~Han,
 {Fermions in loop quantum gravity and resolution of doubling problem}.
 Class. Quant. Grav. \textbf{40}, 205022 (2023).
 {\url{https://doi.org/10.1088/1361-6382/acf26b}}.
 {\href{https://arxiv.org/abs/2212.00933}{{arXiv:2212.00933}}}

\bibitem{Bodendorfer:2011nx}
 N.~Bodendorfer, T.~Thiemann, A.~Thurn,
 {New variables for classical and quantum gravity in all dimensions III.
 Quantum theory}.
 Class. Quant. Grav. \textbf{30}, 045003 (2013).
 {\url{https://doi.org/10.1088/0264-9381/30/4/045003}}.
 {\href{https://arxiv.org/abs/1105.3705}{{arXiv:1105.3705}}}

\bibitem{Han:2013noa}
 Y.~Han, Y.~Ma, X.~Zhang,
 {Connection dynamics for higher dimensional scalar-tensor theories of
 gravity}.
 Mod. Phys. Lett. A \textbf{29}, 1450134 (2014).
 {\url{https://doi.org/10.1142/S021773231450134X}}.
 {\href{https://arxiv.org/abs/1304.0209}{{arXiv:1304.0209}}}

\bibitem{Long:2019nkf}
 G.~Long, C.Y. Lin, Y.~Ma,
 {Coherent intertwiner solution of simplicity constraint in all dimensional
 loop quantum gravity}.
 Phys. Rev. D \textbf{100}, 064065 (2019).
 {\url{https://doi.org/10.1103/PhysRevD.100.064065}}.
 {\href{https://arxiv.org/abs/1906.06534}{{arXiv:1906.06534}}}

\bibitem{Long:2020wuj}
 G.~Long, Y.~Ma,
 {General geometric operators in all dimensional loop quantum gravity}.
 Phys. Rev. D \textbf{101}, 084032 (2020).
 {\url{https://doi.org/10.1103/PhysRevD.101.084032}}.
 {\href{https://arxiv.org/abs/2003.03952}{{arXiv:2003.03952}}}

\bibitem{Long:2020agv}
 G.~Long, Y.~Ma,
 {Polytopes in all dimensional loop quantum gravity}.
 Eur. Phys. J. C \textbf{82}, 41 (2022).
 {\url{https://doi.org/10.1140/epjc/s10052-022-09988-2}}.
 {\href{https://arxiv.org/abs/2009.11196}{{arXiv:2009.11196}}}

\bibitem{Zhang:2011vi}
 X.~Zhang, Y.~Ma,
 {Extension of loop quantum gravity to $f(R)$ theories}.
 Phys. Rev. Lett. \textbf{106}, 171301 (2011).
 {\url{https://doi.org/10.1103/PhysRevLett.106.171301}}.
 {\href{https://arxiv.org/abs/1101.1752}{{arXiv:1101.1752}}}

\bibitem{Zhang:2011qq}
 X.~Zhang, Y.~Ma,
 {Loop quantum $f(R)$ theories}.
 Phys. Rev. D \textbf{84}, 064040 (2011).
 {\url{https://doi.org/10.1103/PhysRevD.84.064040}}.
 {\href{https://arxiv.org/abs/1107.4921}{{arXiv:1107.4921}}}

\bibitem{Zhang:2011vg}
 X.~Zhang, Y.~Ma,
 {Nonperturbative loop quantization of scalar-tensor theories of gravity}.
 Phys. Rev. D \textbf{84}, 104045 (2011).
 {\url{https://doi.org/10.1103/PhysRevD.84.104045}}.
 {\href{https://arxiv.org/abs/1107.5157}{{arXiv:1107.5157}}}

\bibitem{Zhang:2011gn}
 X.D. Zhang, Y.~Ma,
 {Loop quantum Brans-Dicke theory}.
 J. Phys. Conf. Ser. \textbf{360}, 012055 (2012).
 {\url{https://doi.org/10.1088/1742-6596/360/1/012055}}.
 {\href{https://arxiv.org/abs/1111.2215}{{arXiv:1111.2215}}}

\bibitem{Ma:2011aa}
 Y.~Ma,
 {Extension of loop quantum gravity to metric theories beyond general
 relativity}.
 J. Phys. Conf. Ser. \textbf{360}, 012006 (2012).
 {\url{https://doi.org/10.1088/1742-6596/360/1/012006}}.
 {\href{https://arxiv.org/abs/1112.2085}{{arXiv:1112.2085}}}

\bibitem{Chen:2018dqz}
 Q.~Chen, Y.~Ma,
 {Hamiltonian structure and connection-dynamics of Weyl gravity}.
 Phys. Rev. D \textbf{98}, 064009 (2018).
 {\url{https://doi.org/10.1103/PhysRevD.98.064009}}.
 {\href{https://arxiv.org/abs/1803.10807}{{arXiv:1803.10807}}}

\bibitem{Zhang:2020smo}
 X.~Zhang, J.~Yang, Y.~Ma,
 {Canonical loop quantization of the lowest-order projectable Horava gravity}.
 Phys. Rev. D \textbf{102}, 124060 (2020).
 {\url{https://doi.org/10.1103/PhysRevD.102.124060}}.
 {\href{https://arxiv.org/abs/2008.04553}{{arXiv:2008.04553}}}

\bibitem{Zhang:2024khj}
 C.~Zhang, J.~Lewandowski, Y.~Ma, J.~Yang,
 {Black holes and covariance in effective quantum gravity}.
 Phys. Rev. D \textbf{111}, L081504 (2025).
 {\url{https://doi.org/10.1103/PhysRevD.111.L081504}}.
 {\href{https://arxiv.org/abs/2407.10168}{{arXiv:2407.10168}}}

\bibitem{Zhang:2024ney}
 C.~Zhang, J.~Lewandowski, Y.~Ma, J.~Yang,
 {Black holes and covariance in effective quantum gravity: A solution without
 Cauchy horizons}.
 {\href{https://arxiv.org/abs/2412.02487}{{arXiv:2412.02487}}}

\bibitem{Konoplya:2024lch}
 R.A. Konoplya, O.S. Stashko,
 {Probing the effective quantum gravity via quasinormal modes and shadows of
 black holes}.
 {\href{https://arxiv.org/abs/2408.02578}{{arXiv:2408.02578}}}

\bibitem{Liu:2024soc}
 W.~Liu, D.~Wu, J.~Wang,
 {Light rings and shadows of static black holes in effective quantum gravity}.
 Phys. Lett. B \textbf{858}, 139052 (2024).
 {\url{https://doi.org/10.1016/j.physletb.2024.139052}}.
 {\href{https://arxiv.org/abs/2408.05569}{{arXiv:2408.05569}}}

\bibitem{Liu:2024wal}
 H.~Liu, M.Y. Lai, X.Y. Pan, H.~Huang, D.C. Zou,
 {Gravitational lensing effect of black holes in effective quantum gravity}.
 Phys. Rev. D \textbf{110}, 104039 (2024).
 {\url{https://doi.org/10.1103/PhysRevD.110.104039}}.
 {\href{https://arxiv.org/abs/2408.11603}{{arXiv:2408.11603}}}

\bibitem{Malik:2024nhy}
 Z.~Malik,
 {Perturbations and quasinormal modes of the Dirac field in effective quantum
 gravity}.
 {\href{https://arxiv.org/abs/2409.01561}{{arXiv:2409.01561}}}

\bibitem{Heidari:2024bkm}
 N.~Heidari, A.A. Ara\'ujo~Filho, R.C. Pantig, A.~\"Ovg\"un,
 {Absorption, scattering, geodesics, shadows and lensing phenomena of black
 holes in effective quantum gravity}.
 Phys. Dark Univ. \textbf{47}, 101815 (2025).
 {\url{https://doi.org/10.1016/j.dark.2025.101815}}.
 {\href{https://arxiv.org/abs/2410.08246}{{arXiv:2410.08246}}}

\bibitem{Wang:2024iwt}
 Y.~Wang, A.~Vachher, Q.~Wu, T.~Zhu, S.G. Ghosh,
 {Strong gravitational lensing by static black holes in effective quantum
 gravity}.
 {\href{https://arxiv.org/abs/2410.12382}{{arXiv:2410.12382}}}

\bibitem{Skvortsova:2024msa}
 M.~Skvortsova,
 {Quantum corrected black holes: Testing the correspondence between grey-body
 factors and quasinormal modes}.
 {\href{https://arxiv.org/abs/2411.06007}{{arXiv:2411.06007}}}

\bibitem{Ban:2024qsa}
 Z.~Ban, J.~Chen, J.~Yang,
 {Shadows of rotating black holes in effective quantum gravity}.
 {\href{https://arxiv.org/abs/2411.09374}{{arXiv:2411.09374}}}

\bibitem{Du:2024ujg}
 Y.~Du, Y.~Liu, X.~Zhang,
 {Spinning particle dynamics and ISCO in covariant loop quantum gravity}.
 {\href{https://arxiv.org/abs/2411.13316}{{arXiv:2411.13316}}}

\bibitem{Lin:2024beb}
 J.~Lin, X.~Zhang, M.~Bravo-Gaete,
 {Mass inflation and strong cosmic censorship conjecture in covariant quantum
 gravity black hole}.
 {\href{https://arxiv.org/abs/2412.01448}{{arXiv:2412.01448}}}

\bibitem{Konoplya:2025hgp}
 R.A. Konoplya, O.S. Stashko,
 {Transition from regular black holes to wormholes in covariant effective
 quantum gravity: Scattering, quasinormal modes, and Hawking radiation}.
 {\href{https://arxiv.org/abs/2502.05689}{{arXiv:2502.05689}}}

\bibitem{Synge:1966okc}
 J.L. Synge,
 {The escape of photons from gravitationally intense stars}.
 Mon. Not. Roy. Astron. Soc. \textbf{131}, 463 (1966).
 {\url{https://doi.org/10.1093/mnras/131.3.463}}

\bibitem{Hioki:2009na}
 K.~Hioki, K.i. Maeda,
 {Measurement of the Kerr spin parameter by observation of a compact object's
 shadow}.
 Phys. Rev. D \textbf{80}, 024042 (2009).
 {\url{https://doi.org/10.1103/PhysRevD.80.024042}}.
 {\href{https://arxiv.org/abs/0904.3575}{{arXiv:0904.3575}}}

\bibitem{Amarilla:2010zq}
 L.~Amarilla, E.F. Eiroa, G.~Giribet,
 {Null geodesics and shadow of a rotating black hole in extended Chern-Simons
 modified gravity}.
 Phys. Rev. D \textbf{81}, 124045 (2010).
 {\url{https://doi.org/10.1103/PhysRevD.81.124045}}.
 {\href{https://arxiv.org/abs/1005.0607}{{arXiv:1005.0607}}}

\bibitem{Abdujabbarov:2016hnw}
 A.~Abdujabbarov, M.~Amir, B.~Ahmedov, S.G. Ghosh,
 {Shadow of rotating regular black holes}.
 Phys. Rev. D \textbf{93}, 104004 (2016).
 {\url{https://doi.org/10.1103/PhysRevD.93.104004}}.
 {\href{https://arxiv.org/abs/1604.03809}{{arXiv:1604.03809}}}

\bibitem{Tsukamoto:2017fxq}
 N.~Tsukamoto,
 {Black hole shadow in an asymptotically-flat, stationary, and axisymmetric
 spacetime: The Kerr-Newman and rotating regular black holes}.
 Phys. Rev. D \textbf{97}, 064021 (2018).
 {\url{https://doi.org/10.1103/PhysRevD.97.064021}}.
 {\href{https://arxiv.org/abs/1708.07427}{{arXiv:1708.07427}}}

\bibitem{Liu:2020ola}
 C.~Liu, T.~Zhu, Q.~Wu, K.~Jusufi, M.~Jamil, M.~Azreg-A\"\i{}nou, A.~Wang,
 {Shadow and quasinormal modes of a rotating loop quantum black hole}.
 Phys. Rev. D \textbf{101}, 084001 (2020).
 {\url{https://doi.org/10.1103/PhysRevD.101.084001}}.
 {\href{https://arxiv.org/abs/2003.00477}{{arXiv:2003.00477}}}

\bibitem{Kumar:2019ohr}
 R.~Kumar, B.P. Singh, S.G. Ghosh,
 {Shadow and deflection angle of rotating black hole in asymptotically safe
 gravity}.
 Annals Phys. \textbf{420}, 168252 (2020).
 {\url{https://doi.org/10.1016/j.aop.2020.168252}}.
 {\href{https://arxiv.org/abs/1904.07652}{{arXiv:1904.07652}}}

\bibitem{Contreras:2019cmf}
 E.~Contreras, A.~Rinc\'on, G.~Panotopoulos, P.~Bargue\~no, B.~Koch,
 {Black hole shadow of a rotating scale--dependent black hole}.
 Phys. Rev. D \textbf{101}, 064053 (2020).
 {\url{https://doi.org/10.1103/PhysRevD.101.064053}}.
 {\href{https://arxiv.org/abs/1906.06990}{{arXiv:1906.06990}}}

\bibitem{Jusufi:2020odz}
 K.~Jusufi, M.~Azreg-A\"\i{}nou, M.~Jamil, S.W. Wei, Q.~Wu, A.~Wang,
 {Quasinormal modes, quasiperiodic oscillations, and the shadow of rotating
 regular black holes in nonminimally coupled Einstein-Yang-Mills theory}.
 Phys. Rev. D \textbf{103}, 024013 (2021).
 {\url{https://doi.org/10.1103/PhysRevD.103.024013}}.
 {\href{https://arxiv.org/abs/2008.08450}{{arXiv:2008.08450}}}

\bibitem{Jha:2023rem}
 S.K. Jha,
 {Shadow, quasinormal modes, greybody bounds, and Hawking sparsity of loop
 quantum gravity motivated non-rotating black hole}.
 Eur. Phys. J. C \textbf{83}, 952 (2023).
 {\url{https://doi.org/10.1140/epjc/s10052-023-12123-4}}.
 {\href{https://arxiv.org/abs/2310.04759}{{arXiv:2310.04759}}}

\bibitem{Sanchez:2024sdm}
 L.A. S\'anchez,
 {Shadow of a renormalization group improved rotating black hole}.
 Eur. Phys. J. C \textbf{84}, 1056 (2024).
 {\url{https://doi.org/10.1140/epjc/s10052-024-13398-x}}.
 {\href{https://arxiv.org/abs/2408.00226}{{arXiv:2408.00226}}}

\bibitem{Abramowicz:2011xu}
 M.A. Abramowicz, P.C. Fragile,
 {Foundations of black hole accretion disk theory}.
 Living Rev. Rel. \textbf{16}, 1 (2013).
 {\url{https://doi.org/10.12942/lrr-2013-1}}.
 {\href{https://arxiv.org/abs/1104.5499}{{arXiv:1104.5499}}}

\bibitem{Gralla:2019xty}
 S.E. Gralla, D.E. Holz, R.M. Wald,
 {Black hole shadows, photon rings, and lensing rings}.
 Phys. Rev. D \textbf{100}, 024018 (2019).
 {\url{https://doi.org/10.1103/PhysRevD.100.024018}}.
 {\href{https://arxiv.org/abs/1906.00873}{{arXiv:1906.00873}}}

\bibitem{Bambi:2012tg}
 C.~Bambi,
 {A code to compute the emission of thin accretion disks in non-Kerr
 space-times and test the nature of black hole candidates}.
 Astrophys. J. \textbf{761}, 174 (2012).
 {\url{https://doi.org/10.1088/0004-637X/761/2/174}}.
 {\href{https://arxiv.org/abs/1210.5679}{{arXiv:1210.5679}}}

\bibitem{Peng:2020wun}
 J.~Peng, M.~Guo, X.H. Feng,
 {Influence of quantum correction on black hole shadows, photon rings, and
 lensing rings}.
 Chin. Phys. C \textbf{45}, 085103 (2021).
 {\url{https://doi.org/10.1088/1674-1137/ac06bb}}.
 {\href{https://arxiv.org/abs/2008.00657}{{arXiv:2008.00657}}}

\bibitem{Cardoso:2021sip}
 V.~Cardoso, F.~Duque, A.~Foschi,
 {Light ring and the appearance of matter accreted by black holes}.
 Phys. Rev. D \textbf{103}, 104044 (2021).
 {\url{https://doi.org/10.1103/PhysRevD.103.104044}}.
 {\href{https://arxiv.org/abs/2102.07784}{{arXiv:2102.07784}}}

\bibitem{Gan:2021xdl}
 Q.~Gan, P.~Wang, H.~Wu, H.~Yang,
 {Photon ring and observational appearance of a hairy black hole}.
 Phys. Rev. D \textbf{104}, 044049 (2021).
 {\url{https://doi.org/10.1103/PhysRevD.104.044049}}.
 {\href{https://arxiv.org/abs/2105.11770}{{arXiv:2105.11770}}}

\bibitem{Zeng:2021dlj}
 X.X. Zeng, G.P. Li, K.J. He,
 {The shadows and observational appearance of a noncommutative black hole
 surrounded by various profiles of accretions}.
 Nucl. Phys. B \textbf{974}, 115639 (2022).
 {\url{https://doi.org/10.1016/j.nuclphysb.2021.115639}}.
 {\href{https://arxiv.org/abs/2106.14478}{{arXiv:2106.14478}}}

\bibitem{Rosa:2022tfv}
 J.L. Rosa, D.~Rubiera-Garcia,
 {Shadows of boson and Proca stars with thin accretion disks}.
 Phys. Rev. D \textbf{106}, 084004 (2022).
 {\url{https://doi.org/10.1103/PhysRevD.106.084004}}.
 {\href{https://arxiv.org/abs/2204.12949}{{arXiv:2204.12949}}}

\bibitem{Wang:2022yvi}
 H.M. Wang, Z.C. Lin, S.W. Wei,
 {Optical appearance of Einstein-\AE{}ther black hole surrounded by thin
 disk}.
 Nucl. Phys. B \textbf{985}, 116026 (2022).
 {\url{https://doi.org/10.1016/j.nuclphysb.2022.116026}}.
 {\href{https://arxiv.org/abs/2205.13174}{{arXiv:2205.13174}}}

\bibitem{Zeng:2022pvb}
 X.X. Zeng, K.J. He, G.P. Li, E.W. Liang, S.~Guo,
 {QED and accretion flow models effect on optical appearance of
 Euler\textendash{}Heisenberg black holes}.
 Eur. Phys. J. C \textbf{82}, 764 (2022).
 {\url{https://doi.org/10.1140/epjc/s10052-022-10733-y}}.
 {\href{https://arxiv.org/abs/2209.05938}{{arXiv:2209.05938}}}

\bibitem{Yang:2022btw}
 J.~Yang, C.~Zhang, Y.~Ma,
 {Shadow and stability of quantum-corrected black holes}.
 Eur. Phys. J. C \textbf{83}, 619 (2023).
 {\url{https://doi.org/10.1140/epjc/s10052-023-11800-8}}.
 {\href{https://arxiv.org/abs/2211.04263}{{arXiv:2211.04263}}}

\bibitem{Huang:2023ilm}
 Y.X. Huang, S.~Guo, Y.H. Cui, Q.Q. Jiang, K.~Lin,
 {Influence of accretion disk on the optical appearance of the
 Kazakov-Solodukhin black hole}.
 Phys. Rev. D \textbf{107}, 123009 (2023).
 {\url{https://doi.org/10.1103/PhysRevD.107.123009}}.
 {\href{https://arxiv.org/abs/2311.00302}{{arXiv:2311.00302}}}

\bibitem{daSilva:2023jxa}
 L.F.D. da~Silva, F.S.N. Lobo, G.J. Olmo, D.~Rubiera-Garcia,
 {Photon rings as tests for alternative spherically symmetric geometries with
 thin accretion disks}.
 Phys. Rev. D \textbf{108}, 084055 (2023).
 {\url{https://doi.org/10.1103/PhysRevD.108.084055}}.
 {\href{https://arxiv.org/abs/2307.06778}{{arXiv:2307.06778}}}

\bibitem{Wang:2023vcv}
 X.J. Wang, X.M. Kuang, Y.~Meng, B.~Wang, J.P. Wu,
 {Rings and images of Horndeski hairy black hole illuminated by various thin
 accretions}.
 Phys. Rev. D \textbf{107}, 124052 (2023).
 {\url{https://doi.org/10.1103/PhysRevD.107.124052}}.
 {\href{https://arxiv.org/abs/2304.10015}{{arXiv:2304.10015}}}

\bibitem{Bojowald:2015zha}
 M.~Bojowald, S.~Brahma, J.D. Reyes,
 {Covariance in models of loop quantum gravity: Spherical symmetry}.
 Phys. Rev. D \textbf{92}, 045043 (2015).
 {\url{https://doi.org/10.1103/PhysRevD.92.045043}}.
 {\href{https://arxiv.org/abs/1507.00329}{{arXiv:1507.00329}}}

\bibitem{Wald:1984bk}
 R.M. Wald,
 \emph{{General Relativity}}
 (Chicago University Press, Chicago, 1984).
 {\url{https://doi.org/10.7208/chicago/9780226870373.001.0001}}

\bibitem{Liang:2023bk}
 C.~Liang, B.~Zhou,
 \emph{{Differential Geometry and General Relativity: Volume 1}}
 (Springer, Singapore, 2023).
 {\url{https://doi.org/10.1007/978-981-99-0022-0}}

\bibitem{Kumar:2020owy}
 R.~Kumar, S.G. Ghosh,
 {Rotating black holes in $4D$ Einstein-Gauss-Bonnet gravity and its shadow}.
 JCAP \textbf{07}, 053 (2020).
 {\url{https://doi.org/10.1088/1475-7516/2020/07/053}}.
 {\href{https://arxiv.org/abs/2003.08927}{{arXiv:2003.08927}}}

\bibitem{EventHorizonTelescope:2021dqv}
 P.~Kocherlakota {et~al.} (Event Horizon Telescope Collaboration),
 {Constraints on black-hole charges with the 2017 EHT observations of M87*}.
 Phys. Rev. D \textbf{103}, 104047 (2021).
 {\url{https://doi.org/10.1103/PhysRevD.103.104047}}.
 {\href{https://arxiv.org/abs/2105.09343}{{arXiv:2105.09343}}}

\bibitem{Kuang:2022ojj}
 X.M. Kuang, Z.Y. Tang, B.~Wang, A.~Wang,
 {Constraining a modified gravity theory in strong gravitational lensing and
 black hole shadow observations}.
 Phys. Rev. D \textbf{106}, 064012 (2022).
 {\url{https://doi.org/10.1103/PhysRevD.106.064012}}.
 {\href{https://arxiv.org/abs/2206.05878}{{arXiv:2206.05878}}}

\bibitem{Wang:2024lte}
 X.J. Wang, Y.~Meng, X.M. Kuang, K.~Liao,
 {Distinguishing black holes with and without spontaneous scalarization in
 Einstein-scalar-Gauss\textendash{}Bonnet theories via optical features}.
 Eur. Phys. J. C \textbf{84}, 1243 (2024).
 {\url{https://doi.org/10.1140/epjc/s10052-024-13612-w}}.
 {\href{https://arxiv.org/abs/2409.20200}{{arXiv:2409.20200}}}

\bibitem{EventHorizonTelescope:2022wkp}
 K.~Akiyama {et~al.} (Event Horizon Telescope Collaboration),
 {First Sagittarius A* Event Horizon Telescope results. I. The shadow of the
 supermassive black hole in the center of the Milky Way}.
 Astrophys. J. Lett. \textbf{930}, L12 (2022).
 {\url{https://doi.org/10.3847/2041-8213/ac6674}}.
 {\href{https://arxiv.org/abs/2311.08680}{{arXiv:2311.08680}}}

\bibitem{Bambi:2013nla}
 C.~Bambi,
 {Can the supermassive objects at the centers of galaxies be traversable
 wormholes? The first test of strong gravity for mm/sub-mm very long baseline
 interferometry facilities}.
 Phys. Rev. D \textbf{87}, 107501 (2013).
 {\url{https://doi.org/10.1103/PhysRevD.87.107501}}.
 {\href{https://arxiv.org/abs/1304.5691}{{arXiv:1304.5691}}}

\end{thebibliography}
\end{document}